\documentclass[9pt,final, twocolumn]{IEEEtran}
\usepackage{graphicx}
\usepackage{amsthm}
\usepackage{epsfig}
\usepackage{latexsym}
\usepackage{amsfonts}
\usepackage{here}
\usepackage{rawfonts}
\usepackage[latin1]{inputenc}
\usepackage[T1]{fontenc}
\usepackage{calc}
\usepackage{url}
\usepackage{enumerate}
\usepackage{color}
\usepackage[tbtags]{amsmath}
\usepackage{amssymb}
\usepackage{upref}
\usepackage{epic,eepic}
\usepackage{times}
\usepackage{dsfont}
\usepackage{comment}
\usepackage{cite}
\usepackage{bbm} 
\usepackage{multirow}

\usepackage{threeparttable}

\newif\ifrevision
\newcommand{\highlight}[1]{%
\ifrevision
\textcolor{red}{#1}%
\else
#1%
\fi
}

\newtheorem{corollary}{Corollary}
\newtheorem{theorem}{\bf Theorem}

\usepackage{dsfont}

\newlength{\aligntop}
\newlength{\alignbot}
 

\IEEEoverridecommandlockouts

\graphicspath{{./Figures/}}

\usepackage{algorithm}
\usepackage{algorithmic}

\makeatletter
\newcommand\semihuge{\@setfontsize\semihuge{19.3}{25}}
\makeatother

\makeatletter
\newcommand\semismall{\@setfontsize\semihuge{12.4}{15}}
\makeatother

\usepackage{subfigure}

\usepackage{tablefootnote}

\begin{document}

\revisionfalse

\title{Beyond Transmitting Bits: Context, Semantics, and Task-Oriented Communications\vspace*{-0em}
}

\author{Deniz G\"{u}nd\"{u}z,~\IEEEmembership{Fellow,~IEEE}, Zhijin Qin,~\IEEEmembership{Senior Member,~IEEE}, Inaki Estella Aguerri, Harpreet S. Dhillon,~\IEEEmembership{Senior Member,~IEEE}, Zhaohui Yang, Aylin Yener,~\IEEEmembership{Fellow,~IEEE}, Kai Kit Wong,~\IEEEmembership{Fellow,~IEEE}, and Chan-Byoung Chae,~\IEEEmembership{Fellow,~IEEE} 
\thanks{D. G\"{u}nd\"{u}z is with the Department of Electrical and Electronic Engineering, Imperial College London, London SW7 2AZ, U.K. (e-mail: d.gunduz@imperial.ac.uk).}
\thanks{Z. Qin is with the Department of Electronic Engineering, Tsinghua University, Beijing, China (e-mail: qinzhijin@tsinghua.edu.cn).}
\thanks{I. Estella Aguerri is with Amazon, Barcelona, Spain (e-mail: inaki.estella@gmail.com).}
\thanks{H. S. Dhillon is with with Wireless@Virginia Tech, Bradley Department of ECE, Virginia Tech, Blacksburg, 24061, VA, USA (e-mail: hdhillon@vt.edu).}
\thanks{Z. Yang is with Zhejiang Lab, Hangzhou, 31121, China, and also with the College of Information Science and Electronic Engineering, Zhejiang University, Hangzhou, Zhejiang 310027, China, and Zhejiang Provincial Key Lab of Information Processing, Communication and Networking (IPCAN), Hangzhou, Zhejiang, 310007, China (e-mail: yang zhaohui@zju.edu.cn).}
\thanks{A. Yener is with the Department of Electrical and Computer Engineering, The Ohio State University, OH (Email: yener@ece.osu.edu).}
\thanks{K.-K Wong is with the Department of Electronic and Electrical Engineering, University College London, London WC1E 6BT, U.K. (e-mail: kai- kit.wong@ucl.ac.uk).}
\thanks{C.-B. Chae is with School of Integrated Technology, Yonsei University, Seoul, Korea (e-mail: cbchae@yonsei.ac.kr).}
}

\maketitle
%

\begin{abstract}
Communication systems to date primarily aim at reliably communicating bit sequences. Such an approach provides efficient engineering designs that are agnostic to the meanings of the messages or to the goal that the message exchange aims to achieve. Next generation systems, however, can be potentially enriched by folding message semantics and goals of communication into their design. Further, these systems can be made cognizant of the context in which communication exchange takes place, providing avenues for novel design insights. This tutorial summarizes the efforts to date, starting from its early adaptations, semantic-aware and task-oriented communications, covering the foundations, algorithms and potential implementations. The focus is on approaches that utilize information theory to provide the foundations, as well as the significant role of learning in semantics and task-aware communications.
\end{abstract}

\section{Introduction}
Digital communication systems have been conceptualized, designed, and optimized for the main design goal of reliably transmitting bits over noisy communication channels. Shannon's channel coding theory provides the fundamental limit on the rate of this reliable communication, with the crucial design choice that the system design remains oblivious to the underlying meaning of the messages to be conveyed or how they would be utilized at their destination. It is this disassociation from content that allows Shannon's approach to abstract the ``engineering'' problem that is to replicate a digital sequence generated at one point, asymptotically error-free at another. \highlight{While this approach has been tremendously successful in systems where communicating, e.g., voice and data, is the main objective, many emerging applications, from autonomous driving, to healthcare and Internet of Everything, will involve connecting machines that execute (sometimes human-like) tasks. Such cyber-physical systems integrate computation and communication with physical processes where processing units need to communicate and collaborate to monitor and control physical systems through sensor and actuator networks resulting in a networked cyber-physical system. In these applications, the goal of communication is often not to reconstruct the underlying message exactly as in conventional communication networks, but to enable the destination to make the right inference or to take the right decision and action at the right time and within the right context.} Similarly, human-machine interactions will be an important component, where humans will simultaneously interact with multiple devices using text, speech, or image commands, leading to the need for similar interaction capabilities on the devices. These applications motivate the development of ``semantic'' and ``task-oriented'' communication systems. Recent advances in artificial intelligence technologies and their applications have boosted the interest and the potentials of semantics and task-oriented communications, particularly within the context of future wireless communications systems, such as 6G. Integrating semantics into system design, however, is still in its infancy with ample open problems ranging from foundational formulations to development of practical systems. This article provides an introduction to tools and advancements to date in semantic and task-oriented communications with the goal of providing a comprehensive tutorial for communication theorists and practitioners. 
 
 \subsection{Motivation of Semantics and Task-oriented Communications}

There is a growing interest in semantic and goal-oriented communication systems in the recent years. This interest is mainly driven by new verticals that are foreseen to dominate the data traffic in future communication networks. Current communication networks are designed to serve data packets in a reliable and efficient manner without paying attention to the contents of these packets or the impact they would have at the receiver side. However, there is a growing understanding that many of the emerging applications can benefit from going beyond the current paradigm that completely separates the design of the communication networks from the source and destination of the information that flows through the network. The reason behind this trend is twofold: First, the success of many of the emerging applications, such as autonomous driving or smart city/home/factory, relies on massive datasets that enable training large models for various tasks. Hence, supporting and enabling such applications will require carrying significant amounts of traffic due to the transmission of these massive datasets and large models, which can potentially saturate the network capacity. For example, each self-driving car collects terrabytes of data each day from its many sensors, including radar, LIDAR, cameras, and ultrasonic sensors. Such data is often collected by the manufacturers to test and improve their models, generating a huge amount of traffic. Hence, the communication infrastructure cannot simply be an ignorant bit-pipe to enable intelligence at the higher layers, but must incorporate intelligence itself to make sure only the required traffic is transmitted at the necessary time and speed \cite{Gunduz:CM:20, semantic:magazine}. This will require a data-aware communication network design that has the required intelligence to understand the relevance, urgency, and meaning of the data traffic in conjunction with the underlying task. Second, unlike the current content-delivery type traffic, most of the emerging applications require extremely low end-to-end latency. For example, when delivering a video signal to a human, even in the case of live streaming, a certain level of delay is acceptable. However, when this video signal is intended to be used by an autonomous vehicle to detect and avoid potential obstacles or pedestrians on the street, even small delays may not be acceptable. On the other hand, to carry out such a task does not require the vehicle to receive the video sequence with the highest fidelity that is often considered when serving a human receiver. The vehicle is only interested in the content that is relevant for the task at hand.  Again, it is essential to incorporate intelligence into the communication system design to extract and deliver the task-relevant information in the fastest and most reliable manner. 

\subsection{What is Semantics? A Historical Perspective}

Semantics has been studied for centuries in the context of different disciplines, such as philosophy, linguistics, and cognitive sciences, to name a few.  It is a highly complex, and to some extend, controversial topic, which is very difficult, if not impossible, to define in a concise manner that would be widely acceptable. Very briefly, semantics can be defined as the study of ``meaning'', but maybe not surprisingly, it means different things in different areas. `Meaning' itself is an elusive word. 

\highlight{The word semantic originates from the Greek `s\={e}mantic\'{o}s', which means `significant'.} It is closely connected to \textit{semiotics}, the study of \textit{signs}. All communication systems are built upon signs, and a language can be broadly defined as a system of signs and rules. The rules applied to signs are divided into three categories \cite{Morris38}:  1) \textit{Syntax}: studies signs and their relations to other signs; 2) \textit{Semantics}: studies the signs and their relation to the world; and 3) \textit{Pragmatics:} studies the signs and their relations to users. 

In this classification, syntax is only concerned with the signs and their relations, and according to Cherry, ``treats language as a calculus'' \cite{Cherry57}. Semantics is built upon syntax, and its main goal is to understand the relations between signs  and the objects to which they apply, \textit{designata}. According to Chomsky,  syntax is independent of semantics \cite{Chomsky57}. Through his now famous sentence ``Colorless green ideas sleep furiously,'' Chomsky argued that it is possible to construct grammatically consistent but semantically meaningless phrases, hence the separation between syntax and semantics. Pragmatics, on the other hand, is the most general of the three, and considers context of communication; that is, takes into account all the personal and psychological factors (in human communications) when considering the impact of a sign on designata. 



Possibly inspired by the above classification, in his accompanying article to Shannon's Mathematical Theory of Communication in \cite{ShannonWeaver49}, Weaver identified three levels of communication problems, which correspond to the three categories above. According to Weaver, Level A deals with the \textit{technical problem}, and tries to answer the question ``How accurately can the symbols of communication be transmitted?''. Level B instead deals with the \textit{semantic problem}, which asks ``How precisely do the transmitted symbols convey the desired meaning?''. Finally, the third level, Level C, corresponds to the effectiveness problem, and asks ``How effectively does the received meaning affect conduct in the desired way?''. Similarly to syntax in semiotics, the engineering problem can be considered as the syntax in communications, dealing only with the signs used in communication systems, their relations and how they are transmitted over a communication channel. 



Similarly to Chomsky's strict separation between syntax and semantics, Shannon's information theory deals exclusively with the engineering problem, ignoring the meaning of the symbols transmitted. Indeed, in his seminal paper \cite{Shannon}, Shannon explicitly states the following: ``The fundamental problem of communication is that of reproducing at one point either exactly or approximately a message selected at another point. Frequently the messages have \textit{meaning}; that is they refer to or are correlated according to some system with certain physical or conceptual entities. These semantic aspects of communication are irrelevant to the engineering problem. The significant aspect is that the actual message is one selected from a set of possible messages. The system must be designed to operate for each possible selection, not just the one which will actually be chosen since this is unknown at the time of design.'' \highlight{Shannon possibly excluded `meaning' from his theory as he wanted to create a clear mathematical system. He was not alone in this, as many others at the time avoided `meaning' from their studies due to a lack of clear definition and understanding of the word. Morris wrote in \cite{Morris38}: ``The term `meaning' is not included among the basic terms of semiotics. This term, useful enough at the level of everyday analysis, does not have the precision necessary for scientific analysis. Accounts of meaning usually throw a handful of putty at the target of sign phenomena, while a technical semiotic must provide us with words which are sharpened arrows.''}

Despite Shannon's clear indication, many researchers at the time were excited about using Shannon's statistical theory to explain or measure semantics. For example, in \cite{Wiener50}, Wiener wrote ``The amount of meaning can be measured. It turns out that the less probable a message is, the more meaning it carries, which is entirely reasonable from the standpoint of common sense.'' Weaver, on the other hand, although explicitly excluding semantics from Shannon's theory, still argues it has implications for Level B and Level C problems. He writes \cite{ShannonWeaver49} ``[Shannons's theory] although ostensibly applicable only to Level A problems, actually is helpful and suggestive for the level B and C problems.''. And he later adds: ``Thus when one moves to levels B and C, it may prove to be essential to take account of the statistical characteristics of the destination. One can imagine, as an addition to the diagram, another box labeled ``Semantic Receiver'' interposed between the engineering receiver (which changes signals to messages) and the destination. This semantic receiver subjects the message to a second decoding, the demand on this one being that it must match the statistical semantic characteristics of the message to the statistical semantic capacities of the totality of receivers, or of that subset of receivers which constitute the audience one wishes to affect.'' 

Not everybody was in the same opinion as Wiener and Weaver regarding the potential application of Shannon's information theory to semantics. In \cite{BarHillelCarnap53}, Bar-Hillel and Carnap argue that the statistical theory of information conceived by Shannon cannot be applied to study semantics. They also express their dissatisfaction with such attempts: ``Unfortunately, however, it often turned out that impatient scientists in various fields applied the terminology and the theorems of Communication Theory to fields in which the term ``information'' was used, presystematically, in a semantic sense, that is, one involving contents or designata of symbols, or even in a pragmatic sense, that is, one involving the users of these symbols.'' 

Based on Shannon's explicit statement of excluding semantics from his theory, and the ensuing discussion by Carnap and Bar-Hillel, today, many authors argue that the classical Shannon theory cannot handle the semantics related aspects of communication systems. In Shannon's channel coding theorem, the goal is to convey the maximum number of bits through a communication channel in a reliable manner, where ``reliable'' means that the transmitted bit sequence must be reconstructed at the receiver with an arbitrarily low probability of error. Here, each bit is assumed to be equally likely, and what the receiver does with these bits is not relevant for the channel capacity. But, not all information sources generate sequences of equally likely bits. This is obviously not the case in a text in any language. Shannon also looked at such information sources, and showed in his \textit{source coding theorem} \cite{Shannon} that any information source can be compressed into equally likely messages at a rate at least at the entropy of the information source (assuming the information source generates independent symbols from an identical distribution). Even though Shannon explicitly studied the entropy of the English language as an example \cite{Shannon_English}, like channel coding theory, his source coding theory does not deal with the meaning of the words. From the point of Shannon theory, an information source generating messages from the set \{I love you, I miss you, I can't stand you\} is the same as the one generating messages from the set \{1, 2, 3\} as long as the messages come from the same probability distribution.

On the other hand, while Shannon's theory did not deal with the meaning of these messages, it did not ignore the possibility of imperfect reconstructions. Even in his seminal work \cite{Shannon}, which mostly focused on the reliable transmission of sources, Shannon highlighted that the exact transmission of continuous sources would require a channel of infinite capacity, but they can be delivered within a certain fidelity criterion. He laid down the basic ideas of a rate-distortion theory in \cite{Shannon}, although the theory is developed more rigorously only in his later work in 1959  \cite{Shannon_RD}. In \cite{Shannon}, he mentioned various fidelity measures that can be considered when transmitting continuous signals, including mean squared error, frequency weighted mean squared error, and absolute error. Shannon then states: ``The structure of the ear and brain determine implicitly an evaluation, or rather a number of evaluations, appropriate in the case of speech or music transmission. There is, for example, an `intelligibility' criterion in which $\rho(x;y)$ is equal to the relative frequency of incorrectly interpreted words when message $x(t)$ is received as $y(t)$. Although we cannot give an explicit representation of $\rho(x;y)$ in these cases it could, in principle, be determined by sufficient experimentation. Some of its properties follow from well-known experimental results in hearing, e.g., the ear is relatively insensitive to phase and the sensitivity to amplitude and frequency is roughly logarithmic.'' Here, Shannon uses the term ``evaluation'' to refer to different fidelity measures. He clearly makes a reference to reconstruction measures that go beyond recovering a sequence of bits, and allow a certain level of reconstruction error as long as that is within the intelligibility of the receiver, e.g., the ear or the brain. One can further argue that Shannon already hints towards data-driven evaluation of the fidelity of a reconstruction in accordance with the machine learning approaches widely used today.

\highlight{In this paper, our goal is to follow Shannon's approach, in the sense that, we would like to outline a mathematical framework for semantic communication, and highlight how this mathematical framework can be used to design and optimize communication systems and networks. Considering a simple point-to-point communication system, where the goal is to reconstruct the source signal at the receiver end, we treat semantics as a prescribed measure between the source and reconstruction pairs. This measure in general will depend on the underlying goal of the communication, the `significance' of the source signal to be communicated for this goal, and the `fidelity' of the reconstruction in achieving this goal.} We will provide a comprehensive overview of basic Shannon theoretic concepts in relation to semantic and task-oriented communications in this context. While we will provide a brief summary of semantic information theory of Bar-Hillel and Carnap and some of its future extensions as well, resolving the discussion regarding the applicability of Shannon's statistical information theory to study semantics, particularly in the context of linguistics, is out of the scope of this paper. Indeed, our main argument is that if we are given a certain distortion measure on the pairs of transmitted and reconstructed messages, the problem then falls into the realm of information and coding theory. On the other hand, in most cases, particularly for natural languages, characterizing such a distortion measure can be extremely difficult if not impossible. Modern machine learning techniques, particularly those in the area of natural language processing (NLP), can provide such distortion measures, which are becoming increasingly accurate and useful, at least from an engineering standpoint. Therefore, our interpretation and application of semantics to communication systems is closer to Weaver's, where semantics is simply a fidelity measure imposed by the underlying information source, captured by the concept of ``semantic noise'' \cite{ShannonWeaver49}, \highlight{and our focus will be on the theoretical understanding and practical design of communication systems that can extract and exploit such semantic fidelity measures.}

 \subsection{Relevant Surveys and Our Contributions}
Given the increasing interest in semantic- and task-oriented communications, it is not surprising to note that there are quite a few existing surveys and tutorial articles focusing on this general topic already. We discuss them briefly and then list our main contributions. 
The authors in \cite{lan2021semantic} introduced three kinds of semantic communications, human-to-human, human-to-machine, and
machine-to-machine communications. 
The key performance indicators and system design for semantic learning mechanisms over future wireless communications were further pointed out in  \cite{strinati20216g}. 
The goal-oriented signal processing was investigated in \cite{kalfa2021towards}, which included the graph-based semantic language and representation of the semantic information.
Besides, the architectures of semantic communications for artificial intelligence-assisted wireless networks were investigated in \cite{shi2021semantic, kountouris2021semantics, iyer2022survey}.
In \cite{zhang2022toward}, the technical contents and application scenarios were discussed for the intelligent and efficient semantic communication network design. 
In \cite{qin2021semantic}, the authors discussed principles and challenges of semantic communications enabled by deep learning.
Compared with the above works \cite{lan2021semantic,strinati20216g,kalfa2021towards,shi2021semantic,kountouris2021semantics,iyer2022survey,zhang2022toward,qin2021semantic}, the main focus of this paper is to provide a comprehensive introduction to semantic- and task-oriented communications through an information-theoretic viewpoint. In other words, it will be our intention to ground everything discussed in this paper in relevant information-theoretic principles. 


In addition to providing a comprehensive survey of semantic- and task-oriented communication systems, the main ingredients of our paper are listed below. 
\begin{itemize}
\item We first review some of the early efforts in defining a semantic information measure. We point out the differences among existing definitions of semantic entropy, and introduce how knowledge graph based semantics can be applied to and benefit a wide variety of common tasks. 

\item We provide the basic information theoretic concepts for semantic- and task-oriented communications. For instance, we show how the rate-distortion theory can capture the semantic distortion measure and explore the connection between information bottleneck (IB) and  goal-oriented compression. To convey the class information to the receiver, rate-limited remote inference theory is discussed. 

\item Machine learning techniques for  semantic- and task-oriented communications are discussed in two phases separately, i.e., training phase and prediction phase. Various approaches to the remote inference and remote training problems are presented. 

\item We introduce the information theoretic concepts for semantic  and task-oriented transmission over noisy channels from the viewpoint of joint source and channel coding (JSCC). Some practical designs for goal-oriented semantic communication over noisy channels are further provided for text, speech, and image sources.

\item We  discuss the idea of ``timing as semantics'' where the relevance or value of information is in time. Connections of this idea with the related concept of age of information (AoI) are explored by discussing a general real-time remote tracking or reconstruction problem from the rate-distortion viewpoint with a time-sensitive distortion measure which captures the semantic information contained in the timing of source reconstructions. 

\item We present a new effective communication framework, corresponding to the Level C communication problem put forth by Weaver. This results in a context-dependent communication paradigm, where the same message may have a different affect on the receiver depending on the context. 

\end{itemize}

The paper is organized as follows. Some existing definitions for semantic information measures are overviewed in Section II. The information-theoretic foundations of semantic- and task-oriented communications is presented in Section III. Then, in Section IV, relevant machine learning techniques for semantic- and task-oriented communications are introduced in detail. The JSCC approach taking into account the channel effects on semantic information is presented in Section V. To realize semantic communications, some practical designs for goal-oriented semantic communication over noisy channels are  provided in Section VI. As an instance of task-oriented communications, an important class of problems where the metric relates the freshness of information along with connections to semantics is discussed in Section VII. An effective communication framework is presented in Section VIII. Finally, Section IX concludes the paper.



\section{Semantic Information Measures} \label{s:semantic_information}

To extend the engineering approach Shannon proposed, a number of researchers have started to work on a theory of semantic communications soon after Shannon's work. \highlight{One of the desires was to come up with a measure of information, similarly to Shannon's entropy, but one that takes into account the structures within the elements of the set a random variable is defined on, particularly the logical relations. This is in contrast to Shannon's entropy which depends only on the probability distribution of the random variable, but not on the set it is defined on. The concept of semantic entropy was proposed in \cite{carnap1952outline}, and was foreseen to play a significant role in developing a framework that considers semantics.} This notion is then used to quantify semantic information of a source. Until now, a number of other definitions of semantic entropy have been proposed from different perspectives. A commonly-agreed upon notion is yet to emerge however.

\subsection{Semantic Entropy}

\highlight{Semantic entropy is a measure of the average level of "semantic".} In 1952, Carnap and Bar-Hillel~\cite{carnap1952outline} first explicated the concept of semantic entropy of a sentence within a given language system, and provided a way to measure it as follows:
\begin{equation}
    H(s, e) = -\log c(s, e),
\end{equation}
where $c(s, e)$ is the degree of confirmation of sentence $s$ on the evidence $e$, given by
\begin{equation}
    c(s, e) = \frac{m(e,s)}{m(e)}.
\end{equation}
Here, $m(e,s)$ and $m(e)$ represent the \textit{logical probability} of $s$ on $e$ and that of $e$, respectively, \highlight{in which the logical probability is a measure of inductive support or partial entailment. Here, the semantic entropy of a statement is determined by the likelihood of that statement being true `in all possible worlds' \cite{sep-information-semantic}. Statement `s and t' will have higher semantic entropy than statements `s', `t' or `s or t'. For example, the sentence `I slept under the rain' has higher semantic entropy than `I walked under the rain' as the likelihood of the first one being true is smaller. A well-known problem with the Carnap and Bar-Hillel's semantic entropy definition is the fact that it assigns the highest information to the contradictory sentences. This is also knows as the  Bar-Hillel and Carnap Semantic Paradox. }

Different from the logic-based definition, Venhuizen \textit{et al.} \cite{VenhuizenComprehension} derived the semantic entropy based on a language comprehension model in terms of the structure of the world \highlight{(the latent background knowledge)} instead of the logical structure or the probabilistic model of the language, which can be expressed as
\begin{equation}
    \highlight{
    H({\bf{v}}_t)=-\sum\limits_{{{\bf{v}}_M} \in {{\cal V}_{\cal M}}} {p({{\bf{v}}_M}|{{\bf{v}}_t})} \log p({{\bf{v}}_M}|{{\bf{v}}_t}),
    }
\end{equation}
where ${{{\cal V}_{\cal M}}} = \{ {{\bf{v}}_M}|{{\bf{v}}_M}(i) = 1{\rm{ }}\quad {\rm{iff}}\quad{\rm{ }}{M_i} = M\quad{\rm{ and }}\quad M\quad{\rm{\quad is \quad a \quad unique \quad model \quad in }}\quad{\cal M}\}$ is the set of meaning vectors that identify unique models in ${\cal M}$, ${\cal M}$ is the set of models and reflects the probabilistic structure of the world, and \highlight{${p({{\bf{v}}_M}|{{\bf{v}}_t})}$} is the conditional probability of ${{\bf{v}}_M}$ given ${{\bf{v}}_t}$. This comprehension-centric notion of semantic entropy depends on both linguistic experience and world knowledge and quantifies the uncertainty with respect to the whole meaning space.

Apart from the language system, the semantic entropy for intelligent tasks has also been studied. Melamed \cite{melamed1997measuring} proposed a information-theoretic method for measuring the semantic entropy in translation tasks by using translational distributions of words in parallel text corpora. The semantic entropy of each word $w$ is given by 
\highlight{
\begin{equation}
\begin{split}
    H(w)&=H(T|w)+N(w) \\
    &=-\sum\limits_{t \in T}p(t|w) {\log}p(t|w)+p(NULL|w){\log}F(w),
\end{split}
\end{equation}}
where $H(T|w)$ is the translational inconsistency of a source word $w$,  \highlight{which denotes a word that is translated in different ways}, and $T$ represents the set of target words, $N(w)$ denotes the contribution of null links of $w$, \highlight{which indicate the words that do not translate easily from one language to another,} and $F(w)$ is the frequency of $w$. Additionally, for classification tasks, Liu \textit{et al.}~\cite{Liu_Fuzzy:2019} defined the semantic entropy by introducing the membership degree in axiomatic fuzzy set theory. Let ${{\mu _\varsigma }\left( x \right)}$ denote the membership degree of the data sample $x$. The authors first obtained the matching degree \highlight{to measure the probability of the elements belonging to the set,} which characterizes the semantic entropy for data samples in class $C_j,j\in\{1,2,\dots,m\}$ on semantic concept $\varsigma $, as
\begin{equation}
    {D_j}\left( \varsigma  \right) = \frac{{\sum\limits_{x \in {{\cal{X}}_{{C_j}}}}^{} {{\mu _\varsigma }\left( x \right)} }}{{\sum\limits_{x \in {\cal{X}}}^{} {{\mu _\varsigma }\left( x \right)} }},
\end{equation}
where ${{\cal{X}}_{{C_j}}}$ is the set of data for class $C_j$, and ${\cal{X}}$ is the data set of all classes. According to the matching degree, the semantic entropy of class $C_j$ on $\varsigma $ is defined as 
\begin{equation}
    {H_{{C_j}}}\left( \varsigma  \right) =  - {D_j}\left( \varsigma  \right){\log _2}{D_j}\left( \varsigma  \right).
\end{equation}
Further, the semantic entropy of concept $\varsigma $ on ${\cal{X}}$ can be obtained by 
\begin{equation}
    {H}\left( \varsigma  \right) = \sum\limits_{j=1}^{m}{H_{{C_j}}}\left( \varsigma  \right).
\end{equation}
Based on this definition, the optimal semantic descriptions of each class can be obtained and the uncertainty in designing the classifier is minimized. 

In contrast to the aforementioned definitions that are specific to a single task, Chattopadhyay \textit{et al.}~\cite{chattopadhyay2020quantifying} explore an information-theoretic framework to quantify semantic information for any task and any type of source. They define the semantic entropy as the minimum number of semantic queries about data \highlight{$S$}, whose answers are sufficient to predict task $V$, which can be expressed as
\begin{equation}
\highlight{
\begin{split}
    &H_Q(S;V)\buildrel \Delta \over =\mathop {\min }\limits_E \mathbb{E}_S\left[\left| Code_Q^E(S) \right|\right] \\
    &{\rm{s.t.}}\quad P(v|Code_Q^E(s))=P(v|s),\quad \forall s, v,
\end{split}
}
\end{equation}
where $Code_Q^E(s)$ denotes the query vector extracted from \highlight{$S$} with the semantic encoder $E$. From (8), in order to obtain the semantic entropy, one needs to find the optimal semantic encoder, which has the ability to encode $S$ into the minimal representation that can accurately predict the task. Currently, many methods have been utilized for measuring the semantic entropy, such as semantic information pursuit and variational inference. This direction is still in early stages and need to be further investigated.

In summary, there are significant differences among existing definitions of semantic entropy as each of them is based on the properties of its own concerned task. Although the last definition can be extended to different tasks, finding the optimal semantic encoder is as challenging as obtaining semantic entropy. Hence, a unifying definition (as in the case of Shannon entropy) does not exist for semantic entropy, and most of these definitions lack the operational relevance Shannon's entropy enjoys in a large number of engineering problems.

\subsection{Knowledge Graph for Semantic Communications}
\begin{figure}[t]
\begin{centering}
\includegraphics[width=0.38\textwidth]{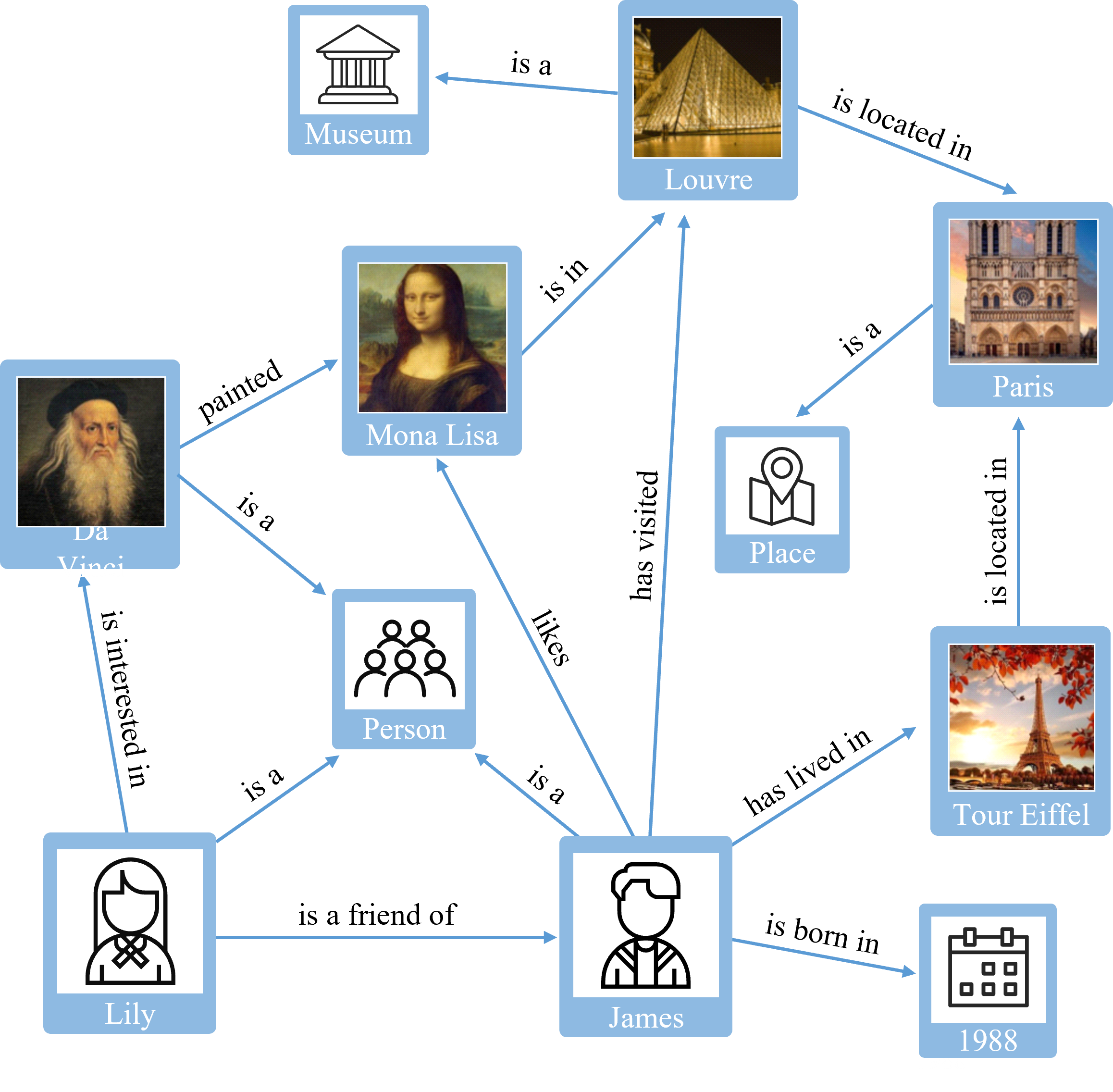}
\par\end{centering}
\caption{Illustration of a knowledge graph.}
\label{KnowledgeGraph}
\end{figure}

\highlight{Knowledge can be defined as the capability to connect available information that facilities understanding and drawing conclusions, and the concept of \textit{knowledge graph} has emerged as an information network to enable such understanding.} Fig. \ref{KnowledgeGraph} presents an example of a knowledge graph, which represents a network of real-world entities, i.e., objects, events, situations, or concepts, and illustrates the relationship between them. This information is usually stored in a graph database and visualized as a graph structure. A knowledge graph is made up of three main components: nodes, edges, and labels. Any object, place, or person can be a node and an edge defines the relationship between the nodes. \textit{Knowledge graph embedding} is to embed components of a knowledge graph including entities and relations into continuous vector spaces, so as to simplify the manipulation while preserving the inherent structure of the knowledge graph~\cite{SurveyKG}. It can benefit a variety of downstream tasks; and hence, has quickly gained attention in the research community. In the following, we review semantic matching models for knowledge graph. Then, we introduce how knowledge graph based semantics can be applied to and benefit from a wide variety of downstream tasks, such as data integration, recommendation systems, and so forth. Subsequently, we present the analysis and framework of knowledge graph based semantics and its applications in semantic communication systems. 

\subsubsection{Semantic Matching Model for Knowledge Graph}
The knowledge graph techniques can be roughly categorized into two groups: translational distance models and semantic matching models. The former use distance-based scoring functions, and the latter similarity-based ones. In the following, we introduce the semantic matching models. In particular, they exploit similarity-based scoring functions and measure plausibility of facts by matching latent semantics of entities and relations embodied in their vector space representations. The authors in~\cite{ThreeWay} associate each entity with a vector to capture its latent semantics. Each relation is represented as a matrix which models pairwise interactions between latent factors. \highlight{Denote a knowledge graph consisting of $n$ entities and $m$ relations, $r$ is the coefficient of rank-$r$ factorization, $A$ represents a $n\times r$ matrix that contains the latent-component representation of the entities and $R_k$, $k\in\left[1,\;2,\;\dots,\;m\right]$, is an asymmetric $r\times r$ matrix that models the interactions of the latent semantics. The factor matrices $A$ and $R_k$ can be computed by solving
the regularized minimization problem,}
\highlight{
\begin{equation}
    \min_{A,\;R_k}\left(f\left(A,\;R_k\right)+g\left(A,\;R_k\right)\right),
\end{equation}}
\highlight{where} 
\highlight{
\begin{equation}
    f\left(A,\;R_k\right)=\frac12\left(\sum_k\left\|{\mathfrak X}_k-AR_kA^T\right\|_F^2\right),
\end{equation}}
\highlight{
\begin{equation}
    {\mathfrak X}_k\approx AR_kA^T,
\end{equation}}
\highlight{and $g\left(\cdot\right)$ is the regularization term,}
\highlight{
\begin{equation}
    g\left(A,\;R_k\right)=\frac12\lambda\left(\left\|A\right\|_F^2+\sum_k\left\|R_k\right\|_F^2\right),
\end{equation}}
\highlight{where $\lambda$ is a constant coefficient.}

In~\cite{SME}, the authors propose the semantic matching energy method, which conducts semantic matching using neural network architectures. It first projects entities and relations to their vector embeddings in the input layer. The relation is then combined with the entity to get the score of a fact. Furthermore, the neural association model has been developed in~\cite{Probabilistic} to conduct semantic matching with a deep architecture.

\subsubsection{Applications of Knowledge Graph Based Semantic}
The knowledge graph embedding is a key technology for solving the problems in knowledge graph. The authors in~\cite{Transms} propose a novel knowledge graph embedding method, which translates and transmits multi-directional semantics: (i) the semantics of head/tail entities and relations to tail/head entities with nonlinear functions, and (ii) the semantics from entities to relations with linear bias vectors. The knowledge graph based semantics has also been employed in data integration~\cite{Integration}, recommendation systems~\cite{Improving}, and real-time ranking~\cite{Compact,Entity,Explicit}. In~\cite{Integration}, the authors devise a semantic data integration approach that exploits keyword and structured search capabilities of Web data sources. The authors in~\cite{Improving} incorporate both word-oriented and entity-oriented knowledge graphs to enhance the data representations, and adopt mutual information maximization to align the word-level and entity-level semantic spaces. In~\cite{Compact}, a novel kind of knowledge representation and mining system has been proposed, which is referred to as the \textit{semantic knowledge graph}. It provides a layer of indirection between each pair of nodes and their corresponding edge, enabling edges to materialize dynamically from underlying corpus statistics. In~\cite{Entity}, an entity-duet neural ranking model has been proposed, which introduces knowledge graphs to neural search systems and represents queries and documents by their words and entity annotations. \highlight{In~\cite{Explicit}, a novel ranking technique that leverages knowledge graphs has been proposed, which addresses the challenge to understand the meaning of research concepts in queries.}

\subsubsection{Analysis and Framework of Knowledge Graph Based Semantic}
In~\cite{Analytics}, the authors introduce the semantic property graph for scalable knowledge graph analytics. To enhance the input data, the authors in~\cite{InputData} propose the framework of relevant knowledge graphs for recommendation and community detection, which improves both accuracy and explainability. In~\cite{Unsupervised}, the authors propose an iterative framework that is based on probabilistic reasoning and semantic embedding. \highlight{The authors in~\cite{Knowledge} represents the knowledge graph by utilizing semantics, which globally extracts semantics in many aspects and then locally assigns a semantic-relevant category in each aspect.}

\subsubsection{Semantic Communication Systems Driven by Knowledge Graph}
In~\cite{Cognitive}, a cognitive semantic communication framework is proposed by exploiting knowledge graph. Moreover, a simple and interpretable solution for semantic information detection is developed by exploiting triplets as semantic symbols. It also allows the receiver to correct errors occurring at the symbolic level. Furthermore, the pre-trained model is fine-tuned to recover semantic information, which overcomes the drawback that a fixed bit length coding is utilized to encode sentences of different lengths.


\section{Information Theoretic Foundations of Semantic- and Task-Oriented Communications}\label{s:semantic_compression}

As we have mentioned, despite the many efforts to define a semantic information measure, none of the aforementioned attempts have so far resulted in a widely accepted definition, or had an impact on operational performance similarly to Shannon's information theory had on communication systems. Therefore, we take a statistical approach to semantics in this paper, and either treat it as a given distortion measure, or consider data-driven approaches to acquire it. \highlight{Our goal is to abstract out the linguistic aspects of semantic communications, which is out of the scope of this paper, and instead focus on the communication challenges.}

In the current section, we will provide some of the basic concepts in Shannon's statistical information theory, and how they can be used to study semantics in emerging communication systems. In particular, we will first review rate-distortion theory, and its characterization in various settings and under different distortion measures, and how they can capture semantic or goal-oriented communications. In this section, we will focus on rate-limited error-free communications; that is, we will mainly treat the semantic/ task-oriented compression problem. Semantic- and task-oriented transmission over noisy channels will be considered in Section \ref{s:JSCC}.

\subsection{Rate-Distortion Theory}\label{ss:rate_distortion_theory}

In \cite{Shannon_RD}, Shannon expands the ideas put forth in his seminal work by formally defining the problem of lossy source transmission, which laid down the principles of \textit{rate-distortion theory}.  Note that, when compressing a single source sample we have a \textit{quantization} problem, which is equivalent to analog-to-digital conversion when the source samples come from a continuous alphabet. This can also be treated as a clustering problem with a specified fidelity criterion. Shannon showed that, similarly to the channel coding theorem, it is more efficient, in terms of bits per source sample, if many independent samples of the same source distribution can be compressed together.   

Let $s^m \in \mathcal{S}^m$ denote $m$ independent source samples distributed according to $p_S(s)$, i.e., $p(s^m) = \prod_{i=1}^m p_S(s_i)$. Let $\hat{s}^m \in \mathcal{\hat{S}}^m$ denote the reconstruction at the receiver, where $\mathcal{\hat{S}}$ is the reconstruction alphabet, not necessarily same as the source alphabet. The goal is to minimize the distortion between $s^m$ and $\hat{s}^m$ under some given distortion (fidelity) measure, $d: \mathcal{S}^m \times \mathcal{\hat{S}}^m \rightarrow [0, \infty)$, which assigns a certain distortion, or equivalently, a quality metric, for every pair of source and reconstruction sequences.   Shannon considered single-letter additive distortion measures, that is, $d(s^m, \hat{s}^m) = 1/m \sum_{i=1}^m d(s_i, \hat{s}_i)$ for a 
single-letter distortion measure $d: \mathcal{S} \times \mathcal{\hat{S}} \rightarrow [0, \infty)$. 

The goal in lossy source coding is to represent the source sequence with as few bits as possible, measured in \textit{bits per source sample (bpss)}, while guaranteeing a certain average distortion level. A $(m,2^{mR})$ lossy source code consists of an \textit{encoder}-\textit{decoder} pair, where the encoder $f^{(m)}: \mathcal{S}^m \rightarrow [2^{mR}]$ maps each $m$-length source sequence $s^m \in \mathcal{S}^m$ to an index  $w(s^m) \in  [2^{mR}]$, and the decoder $g^{(m)}: [2^{mR}] \rightarrow \mathcal{\hat{S}}^m $ maps each index $w \in  [2^{mR}]$, to an estimated reconstruction sequence $\hat{s}^m(w) \in \mathcal{\hat{S}}^m$, where for $a \in \mathds{R}$, we have defined $[a]\triangleq \{1, \ldots \lfloor a \rfloor\}$. The collection of all reconstruction sequences $\{\hat{s}^m(1), \ldots, \hat{s}^m(2^{(mR)})\}$ forms the codebook, which is assumed to be shared between the encoder and decoder in advance. 

The average distortion of a $(m,2^{mR})$ code is given by
\begin{equation}
	\mathds{E}[d(S^m, \hat{S}^m)] = \sum_{s^m \in \mathcal{S}^m} p(s^m) d(s^m, \hat{s}^m(w(s^m))),
\end{equation}
where the expectation is taken over the source distribution. 

For a given source distribution $p_S(s)$ and distortion measure $d(s, \hat{s})$, we say that a rate-distortion pair $(R,D)$ is \textit{achievable} if there exist a sequence of $(m,2^{mR})$ codes that satisfy
\begin{equation}
\limsup_{m \rightarrow \infty} \mathds{E}[d(S^m, \hat{S}^m)] \leq D.
\end{equation}
The rate-distortion function $R(D)$ of source $S$ under the single-letter distortion measure $d(\cdot, \cdot)$ is defined as the infimum of rates $R$ for which $(R,D)$ is achievable. 

For such single-letter additive distortion measures, Shannon provided a single-letter characterization of the optimal rate-distortion function. 

\begin{theorem}(Shannon's Lossy Source Coding Theorem, \cite{Shannon_RD})  The rate-distortion function for source $S$ and distortion measure $d(\cdot, \cdot)$ is given by
\begin{equation}
	R(D) = \min_{p(\hat{s}|s): \mathrm{E}[d(S, \hat{S}) \leq D]} I(S;\hat{S}).
\end{equation}
\end{theorem}

We can argue that Shannon's rate distortion function quantifies the communication rate required to convey sample-level semantic information when many source samples can be compressed jointly. Here, our assumption is that the semantic relevance of reconstructing source sample $s$ as $\hat{s}$ at the receiver is quantified by the prescribed distortion measure $d(s,\hat{s})$. In the context of text, this may refer to reconstructing a certain sentence in a manner that preserves  its core meaning. In the context of music or image, it may mean, as argued by Shannon, preserving the intelligibility of the source signal, for example by  conveying only the most audible or most distinguishable frequencies, which is the approach followed by modern audio and image compression standards. 

We note, however, that in most cases we apply compression algorithms on a single sample, e.g., a single image, a single video sequence, or a single text file. In these cases, samples correspond to pixels or patches in an image, frames in a video, and letters or words in the text. Such samples are often highly correlated, and additive distortion measures may not preserve the semantics of the overall content. For example, preserving word level similarity, e.g., replacing words with their more common synonyms in its reconstruction, or generating an image with a low pixel-level mean squared error (MSE) may not lead to a good quality semantic compression. For text, semantic compression requires a much deeper understanding of the semantics of the underlying language. For images, alternative quality measures have been proposed that would provide a better image level semantic reconstruction. For example,  structural similarity index measure (SSIM) or multi-scale SSIM (MS-SSIM) \cite{wang_multiscale_2003} have been introduced to measure the perceived quality of images and videos by incorporating luminance masking and contrast masking terms into the distortion measure, providing perceptually more satisfactory reconstructions \cite{Huang:TCSVT:10}. \highlight{However, it has been shown in \cite{Patel:perception} that the reported improvements in the MS-SSIM performance may be misleading for the perceptual quality of the reconstructed images, and a higher MS-SSIM metric does not necessarily lead to better perceptual quality. An alternative perception-based quality metric, called learned perceptual image patch similarity (LPIPS) is proposed in \cite{Zhang:CVPR:18}, which trains a convolutional neural network (CNN) on user judgments on distorted images to compute this metric. Moreover, SSIM/MS-SSIM do not provide adaptivity to regions of interest in an image. In \cite{Zhicheng:IVC:11, Khanna:PerMIn:15}, saliency-based attention prediction is used to detect regions-of-interest in image and video signals, which are then used for adaptive bit allocation.}

\subsection{Rate-Distortion-Perception Trade-off}\label{ss:rate_distortion_perception}

Recently, it has been observed that perceived quality of reconstructed images can be improved using generative adversarial networks (GANs) \cite{NIPS2014_5ca3e9b1}. GANs train a \textit{discriminator} that tries to distinguish the reconstructed image from the images in the dataset, forcing the decoder to generate realistic looking images. GANs employ distance measures between the reconstructed image and the statistics of the images in the training dataset \cite{NEURIPS2018_801fd8c2, Agustsson_2019_ICCV}, such as the Jensen-Shannon divergence, the Wasserstein distance \cite{pmlr-v70-arjovsky17a}, or an f-divergence \cite{NIPS2016_cedebb6e}. By taking the perceptual quality into account when reconstructing a  compressed image, we aim not only to reproduce the original image with the highest fidelity, but also to reconstruct a more natural realistic image, which preserves the semantics of the underlying distribution.

\begin{figure}[t]
     \centering
       \includegraphics[width=\linewidth]{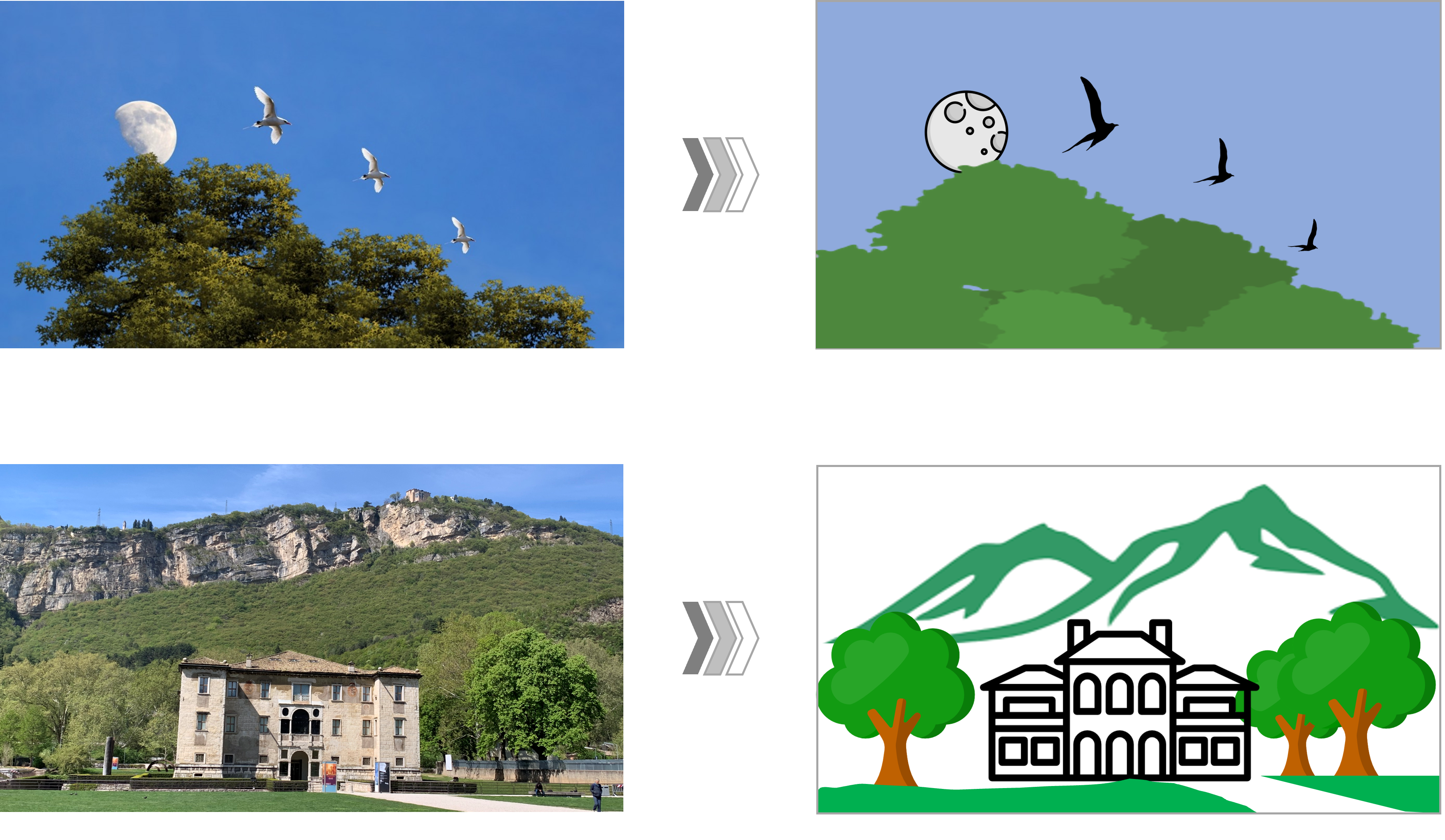}
     \caption{Examples of semantic image compression.}
     \label{f:semantic_image_comp}
 \end{figure}

Semantic information can be used to provide a more efficient compression algorithm, or to achieve a better quality reconstruction \cite{Agustsson_2019_ICCV, Akbari:ICASSP:19, Kushal:SM:20}.  See, for example, the images in Fig. \ref{f:semantic_image_comp}. By only extracting and transmitting semantic information, e.g., the objects in the original image and their general layout, the output image can be reconstructed by simply including a generic representative of each of the  objects in the image. Hence, it is possible to convey the image at a level to enable semantic reasoning about the image, albeit not reliably at the pixel level. 

\highlight{
In \cite{Agustsson_2019_ICCV}, the authors used GANs to push the limits of image compression in very low bit-rates by synthesizing image content, such as facades of buildings, using a reference image database. A selective generative compression method is proposed, which generates parts of the image from a semantic label map, which can be obtained using a semantic segmentation network \cite{Zhao_2017_CVPR, Fu_2019_CVPR}. These parts of the image are fully synthesized rather than being reconstructed based on the original image. This allows the receiver to generate images that resemble the source image semantically, although they may not match perfectly in details, providing visually pleasing reconstructions even at very low bit-rates. The rest of the image is generated using a conditional generative adversarial network (cGAN) \cite{DBLP:journals/corr/MirzaO14}. 
}

A deep semantic segmentation-based layered image compression scheme is proposed in \cite{Akbari:ICASSP:19}, where the semantic map of the input image is used to synthesize the image, while a compact representation and a residual are further encoded as enhancement layers. It is shown that this semantic-based compression approach outperforms BPG and other standard codecs in both PSNR and MS-SSIM metrics. We also add that, including the segmentation map in the bit-stream can further facilitate other downstream tasks such as image search or compression and manipulation of individual image segments.

Another semantic-based image processing approach uses  scene graphs to extract not only objects within the image, but also their relationships \cite{Johnson:CVPR}. A scene graph is a directed graph data structure consisting of the objects and their attributes as vertices, and the relations between the objects as edges. Scene graph generation typically follows three steps: i) Feature extraction, which is responsible for identifying the objects in the image; ii) Contextualization, which established contextual information between entities, and finally iii) Graph construction and inference, which generates the final scene graph using the contextual information, and carry out desired inference tasks on the graph \cite{Chang:scene}. Scene graphs are powerful tools that can encode images \cite{Johnson:CVPR} or videos \cite{Wang:2020} using abstract semantic elements. 

More formal definitions of the perceptual quality of an output image has also been considered, defined as `the extent to which it is perceived as a valid (natural) sample' \cite{pmlr-v97-blau19a}. We note that, while the distortion of image $\hat{s}^m$ is defined with respect to the source image $s^m$, perceptual quality is defined as a property only of the reconstruction. Since the perception does not necessarily depend on the input image, a generative model can generate an image that would exhibit the same statistical properties of the images in the dataset, achieving perfect perception quality, even in the extreme case of zero-compression rate. As the compression rate increases, we expect the reconstruction to better resemble the particular source image. This idea has been identified as the \textit{distribution-preserving image compression} in \cite{Tschannen:NEURIPS:2018}.

\highlight{The perceptual quality of an image is often defined as the divergence between the distribution of the reconstruction and the statistics  of natural images \cite{Wang:TIP:06}. If one can make this divergence vanish, it  means that the reconstruction is indistinguishable from real data samples; however, this does not guarantee that the distortion between the particular input sample and its reconstruction at the decoder is small. It is shown in \cite{pmlr-v97-blau19a} that in general there is a tension between the distortion that can be achieved and the divergence to the real data distribution, which is formalized as the \textit{rate-distortion-perception trade-off}. An information theoretic formulation of the rate-distortion-perception trade-off is presented in \cite{Theis:ICLR:21} allowing stochastic and variable-length codes.}

\highlight{Consider the source $S^m$ to be compressed as before, and the distortion function $d(\cdot, \cdot)$, where $d(s,\hat{s})= 0$ iff $s=\hat{s}$. We further assume the presence of common randomness between the encoder and the decoder. Hence, we define a $(m,2^{mR}, 2^{mR_c})$ code as a pair of an encoding function 
\[
f^{(m)}: \mathcal{S}^m \times [2^{mR_c}] \rightarrow [2^{mR}],
\] 
and a decoding function 
\[
g^{(m)}: [2^{mR_c}] \times [2^{mR}] \rightarrow \hat{\mathcal{S}}^m.
\]
}

\highlight{
We then say that a tuple $(R, R_c, D, P)$ is \textit{achievable}, if there exist a sequence of $(m, 2^{mR}, 2^{mR_c})$ codes, such that 
\[
\limsup_{m\rightarrow \infty} \mathds{E}[d(S^m, \hat{S}^m)] \leq D
\]
and
\[
\limsup_{m\rightarrow \infty} \delta(P_{S^m}, P_{\hat{S}^m}) \leq P,
\]
where $\hat{S}^m = g^{(m)}(f^{(m)}(S^m, J), J)$ and $J$ is the common randomness uniformly distributed over $[2^{(R_c}]$ and independent of $S$. Here, $\delta(\cdot, \cdot)$ is an appropriate measure of similarity between distributions, e.g., total variation distance or KL divergence. The goal, as before, is to characterize the set of achievable $(R, R_c, D, P)$ tuples for a given source distribution $P_S$, distortion measure $d$, and divergence $P$.
}

\highlight{
A one-shot version of this problem is considered in \cite{Theis:ICLR:21}, where it is first argued that the perception constraint can simply be interpreted as yet another distortion measure on the joint distribution of the source and its reconstruction. Then, using results from \cite{Li:TIT:18}, an operational rate-distortion-perception function (RDPF) is defined, and associate lower and upper bounds are proven for variable-length coding with common randomness. RDPF is characterized for a Gaussian source under MSE distortion measure, and Wasserstein-2 or Kullback-Leibler divergence for the perception loss in \cite{Zhang:NeurIPS:21}. In \cite{Zhang:NeurIPS:21}, authors focused on the universal version of the rate-distortion-perception trade-off, and showed that fixing a good representation map and only varying the decoder may be sufficient to achieve multiple points on this trade-off. 
}

\highlight{
In conventional rate-distortion theory, it can be proven that deterministic encoders are sufficient to achieve the optimal rate-distortion performance. This simplifies both the analysis and the implementation of rate-distortion optimal codes. However, in the case of rate-distortion-perception trade-off, it has been shown in \cite{Theis:ICLR:21b} that stochastic encoders can be strictly better than their deterministic counterparts. The authors show that stochastic encoders can be particularly beneficial when perfect perceptual reconstruction is desired; that is, when $P \rightarrow 0$. This points to the requirement of some common randomness.
}

\highlight{
The characterization of the optimal rate-distortion-perception trade-off is given in \cite{Wagner:RDPT:22} under the total variation distance as the perception measure and for the perfect perception case, i.e., $P=0$. }

\highlight{
\begin{theorem}[\cite{Wagner:RDPT:22}]\label{t:tradeoff}
For a finite-alphabet source $S$, the tuple $(R, R_c,D, 0)$ is achievable iff there exist $(U,\hat{S})$ such that
\begin{eqnarray}
    P_S = P_{\hat{S}}, \\
    S - U - \hat{S}, \\
    R \geq I(S;U), \\
    R + R_c \geq I(\hat{S}; U), \\
    \Delta \geq E[D(S, \hat{S})].
\end{eqnarray}
\end{theorem}
}

\highlight{
This characterization explicitly shows the impact of the amount of available common randomness on the achievable trade-off. When sufficient common randomness is available, this results boils down to the one in \cite{Li:arXiv:11}. On the other hand, when there is no common randomness, the trade-off boils down to the following region.
}

\highlight{
\begin{corollary}
For a finite-alphabet source $S$, the tuple $(R, 0, D, 0)$ is achievable iff there exist $(U,\hat{S})$ such that
\begin{eqnarray}
    P_S = P_{\hat{S}}, \\
    S - U - \hat{S}, \\
    R \geq \max\{I(S;U), I(U;\hat{S})\}, \\
    \Delta \geq E[D(S, \hat{S})].
\end{eqnarray}
\end{corollary}
}

\highlight{
In \cite{Wagner:RDPT:22}, the optimal rate-distortion trade-off for perfect perception is explicitly characterized for a Gaussian source and MSE distortion measure. An interesting observation based on this result is that, at small distortions, requiring perfect perception at the decoder incurs no rate penalty as long as sufficient common randomness is available. 
}

In the context of video coding, semantic-based compression has long been considered for very low bit-rate video compression \cite{Manoranjan}. These include motion-compensated compression, where motion vectors of pixels between two consecutive images or optical flow vectors \cite{HORN1981185} are encoded and transmitted. Alternative \textit{object-based compression} methods have also been considered in the literature for a very long time \cite{OSTERMANN1994143, Talluri:TCSVT:97}. In object-based compression, each moving object in a video signal is separated from the stationary background and are conveyed to the decoder by describing their shape, motion, and content using an object-dependent parameter coding. Using this coded parameter set, each image is then synthesized at the decoder by model-based image synthesis.  Although such an object-based compression approach was standardised as part of MPEG4 in the late 90s, it has not been widely adopted due to the lack of fast and reliable object and motion segmentation  techniques. This approach is regaining interest in recent years due to the rapid advances in deep learning based segmentation techniques \cite{Duan:TIP:20, visapp21}.  

In general, quantifying the semantic distortion measure for a particular information source is a formidable task. There have been many studies to understand and model semantics particularly in the context of text and natural language processing. We will go over some of these efforts in Section \ref{s:JSCC} in the context of semantic/goal-oriented transmission. 

	\begin{figure}
		\centering
		\includegraphics[width=\linewidth]{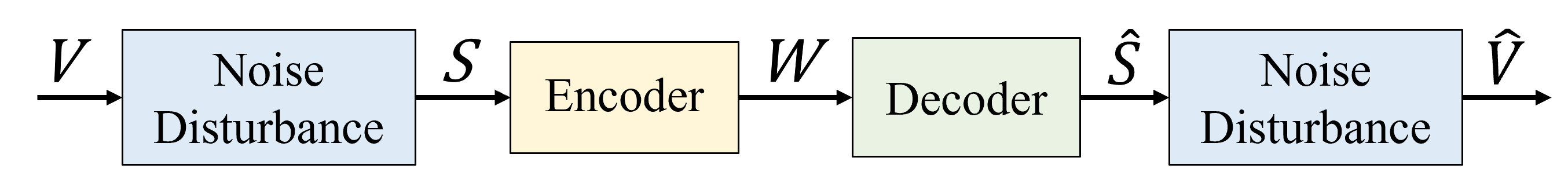}
		\caption{Remote source-compression problem where the encoder observes a noisy version $S$ of the remote source $V$ through a random disturbance $p_{S|V}$, and the  reconstruction $\hat{S}$ goes through another random transformation $p_{\hat{V}|\hat{S}}$. The goal is to guarantee a certain fidelity requirement between $V$ and $\hat{V}$. }
		\label{f:remoteRD}
	\end{figure}

\subsection{Goal Oriented Compression}\label{ss:goal_oriented}

In the conventional rate-distortion framework overviewed above,  the goal is to reconstruct the source sequence $s^m$ within a desired fidelity constraint. This framework applies to most delivery scenarios, such as the transmission of an image, video or audio source over a rate-limited channel, where the goal is to recover the original content with the highest fidelity. However, in many emerging applications, particularly those involving machine-type communications, the receiver may not be interested in the source sequence, but only in a certain feature of it. For example, rather than reconstructing an image, the receiver may be interested in certain statistical aspects of the image, or the presence or absence of certain objects or persons in the image. This can model a goal-oriented compression scenario, where reconstructing the desired feature can represent the goal.  

The desired feature in this context can be modelled as a correlated random variable $V\in \mathcal{V}$ that follows a joint distribution $p(v,s)$ with the source sample $S$. In this scenario, we can instead impose a distortion constraint on the feature and its reconstruction at the decoder, for a prescribed distortion measure $d(v, \hat{v}) < \infty$ for $(v,\hat{v}) \in \mathcal{V} \times \mathcal{\hat{V}}$. This problem can be interpreted as a \textit{remote source coding problem}, which was originally studied by Dobrushin and Tsybakov in  \cite{Dobrushin}. In addition to the noisy observation of the features at the encoder, they also considered another random transformation at the output of the decoder, as illustrated in Fig. \ref{f:remoteRD}. They showed that this generalization of the Shannon's rate-distortion problem can be reduced to Shannon's original formulation by appropriately transforming the distortion measure. Consider any pair of encoding function $f$ and decoding function $g$. The average end-to-end distortion achieved by this encoder-decoder pair is given by 
\begin{equation}
	D(f, g) = \sum_{v, s, \hat{s}, \hat{v}} p_V(v) p_{S|V}(s|v)p^{(f,g)}_{\hat{S}|S}(\hat{s}|s) p_{\hat{V}|\hat{S}}(\hat{v}|\hat{s})  d(v, \hat{v}).
\end{equation}

From the perspective of the encoder, it observes a source $S$ with marginal distribution $p_S(s) = \sum_{v\in\mathcal{V}} p_V(v) p_{S|V}(s|v)$. Let us now consider the modified distortion measure
\begin{equation}
	\hat{d}(s,\hat{s}) = \frac{1}{p_S(s)} \sum_{v, \hat{v}} p_V(v) p_{S|V}(s|v) p_{\hat{V}|\hat{S}}(\hat{v}|\hat{s})  d(v, \hat{v}).
\end{equation}
Using this new distortion measure, we can rewrite the end-to-end distortion as 
\begin{equation}
	D(f, g) = \sum_{s, \hat{s}} p_S(S) p^{(f,g)}_{\hat{S}|S}(\hat{s}|s) ) \hat{d}(s, \hat{s}).\label{eq:new_distortion}
\end{equation}
Therefore, the problem of minimizing the average end-to-end distortion for the remote source coding problem can be reduced to the classical source coding problem for a source with marginal distribution $p_S$ under the modified distortion measure $\hat{d}(\cdot, \cdot)$. 

The following equivalence can be generalized to the standard block coding setting. Assume now that we observe a sequence $s^m$ and want to  reconstruct the corresponding feature vector $v^m$, where $(v_i, s_i)$ are i.i.d. samples from the joint distribution $p_{V,S}$. Then, the corresponding \textit{remote rate-distortion function} can be characterized in a single letter form as given in the next theorem. 

\begin{theorem}(Remote Rate-Distortion Function, \cite{Dobrushin})  The remote rate-distortion function for source $V$ based on its observation $S$ following joint distribution $p_{V,S}(v,s)$ and distortion measure $d(\cdot, \cdot)$ is given by
\begin{equation}
	R^{\mathrm{remote}}(D) = \min_{p(\hat{s}|s): \mathrm{E}[\hat{d}(S, \hat{S}) \leq D]} I(S;\hat{S}).
\end{equation}
\end{theorem}

Recently, the remote rate-distortion interpretation of semantic compression is considered in \cite{Liu:ISIT:21}, where the decoder wants to reconstruct both the feature vector $v^m$ and the source vector $s^m$, under different distortion measures. With the above reduction, one can see that this problem trivially reduces to the Shannon rate-distortion problem with two  distortion measures. Further characteristics of this rate-distortion function for a Hamming distortion measure is studied in \cite{Stavrou2022}. In \cite{Wolf:TIT:70}, it is shown that the optimal transmission scheme for the general model in Fig. \ref{f:remoteRD} under the squared error distortion measure can be divided into two steps: the encoder first estimates the feature variable $v$ conditioned on $s$, and then conveys the estimated value to the decoder. 

	\begin{figure}
		\centering
		\includegraphics[width=\linewidth]{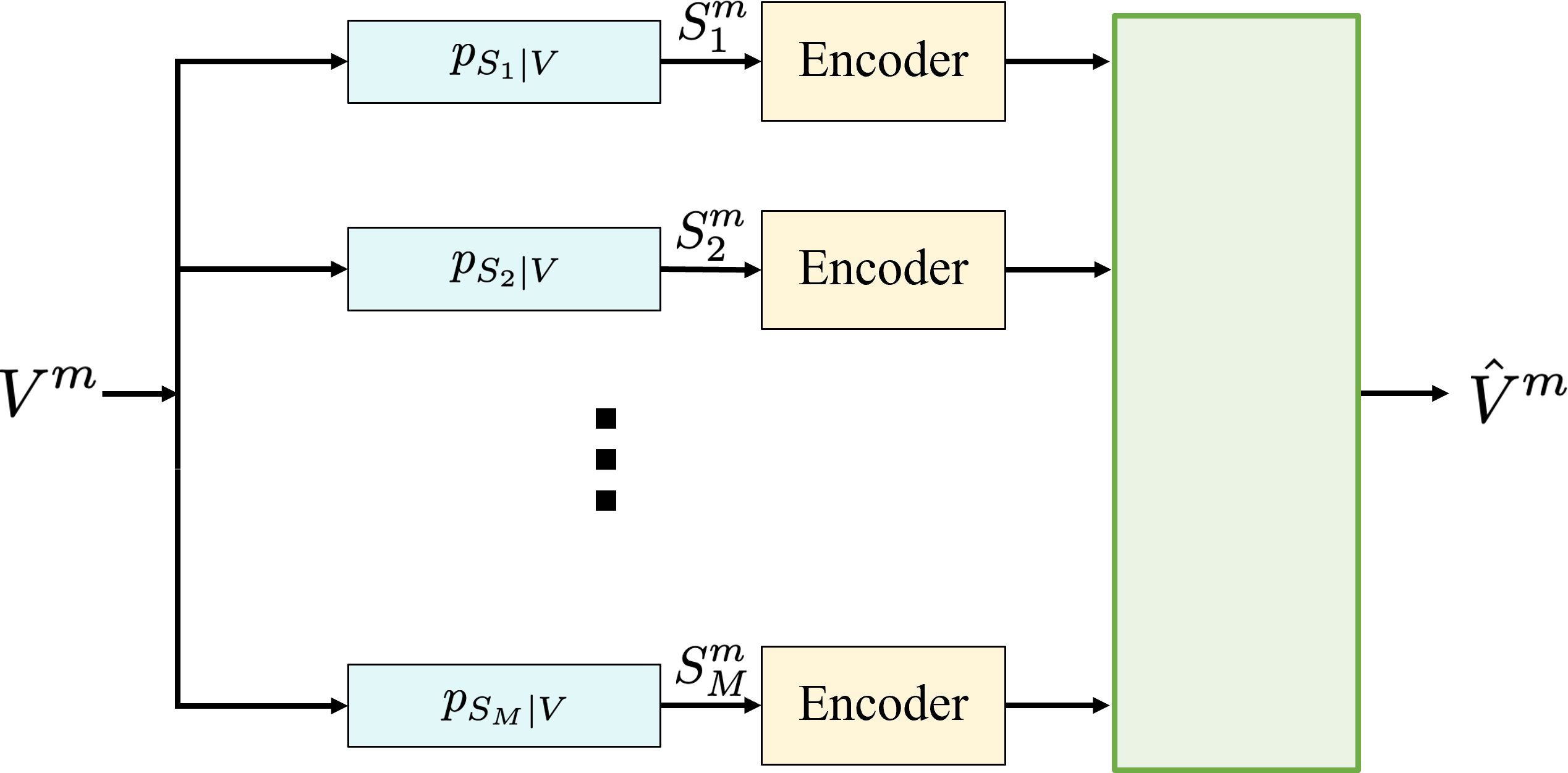}
		\caption{The CEO problem.}
		\label{f:CEO}
	\end{figure}

A natural extension of the above remote rate-distortion function involves multiple terminals, each observing a different noisy version of the underlying latent source $V$ (see Fig. \ref{f:CEO} for an illustration). This is known as the ``CEO problem'' in the literature following \cite{Berger:TIT:96}. In this setting, a chief executive officer (CEO) is interested in estimating an underlying source sequence, $V^n$. She sends $M$ agents to observe independently corrupted versions of the source sequence, where the observations $S_i^m$ of the $i$-th agent are generated through the conditional distribution $p_{S_i|V}$ in an i.i.d. manner. The agents cannot communicate among each other, and each one only has a rate-constrained channel to the CEO. For a given sum rate constraint, what is the minimum average distortion the CEO can estimate $S^n$ under a given fidelity measure $d(\cdot, \cdot) < \infty$?  The special case of Gaussian source and noise statistics with squared error distortion is called the quadratic Gaussian CEO problem \cite{Viswanathan:TIT:97}. The problem remains open in the general case, while the rate region was characterized for the Gaussian case in \cite{Oohama:TIT:05}, for logarithmic loss distortion for discrete sources in \cite{CW14} and vector Gaussian sources in \cite{UEZb}.

\subsection{Context as Side Information}\label{ss:context_SI}
When the communication of an underlying source signal is considered, additional information that is correlated with this desired source variable available to the transmitter and the receiver can be leveraged to reduce the rate requirements. Consider for example a surveillance camera in a house recording a video and forwarding the recording to a remote node which aims at detecting the presence of intruders. Depending on the hour of the day, the illumination of the image will be different. The context in which the information is being obtained, e.g., the time, or weather conditions, could be exploited to improve the video compression quality. Images of the same scene recorded by other cameras can also serve as context information. 

From an information theoretical perspective, contextual information can be modelled as side information. The problem of lossy source coding when  common correlated side information is available both at the encoder and the decoder was studied by Gray in \cite{G72}. By characterizing the rate-distortion function for this problem, it was shown that the rate required to achieve a prescribed distortion could be reduced by exploiting the side information available.  Interestingly, the rate required to achieve a particular distortion is also reduced when the correlated side information is only available at the decoder, and the encoder has to compress without knowledge of the realization of the side information available, as long as its distribution is known. This problem was studied in \cite{WZ76}, and the corresponding single-letter rate-distortion function is referred to as the Wyner-Ziv rate-distortion function. 

Let $T^m$ denote the side information sequence observed by the receiver that is correlated with the source $S^m$. In particular, let $(S_i,T_i)$ be i.i.d. samples jointly distributed with $p(S,T)$. 
The Wyner-Ziv rate distortion function for source coding problem with side information at the decoder is characterized as given in the following theorem.

\begin{theorem}(Lossy Compression with Side Information at the Decoder, \cite{WZ76}) The rate-distortion function for source $S$ with side information $T$ available only at the decoder following joint distribution $p_{S,T}(s,t)$ and distortion measure $d(\cdot, \cdot)$ is given by
\begin{equation}
	R^{\mathrm{WZ}}_{S|T}(D) = \min I(S;U|T),
\end{equation}
where the minimum is over conditional distributions $p_{U|S}$ with $|\mathcal{U}| \leq |\mathcal{S}|+1$ and functions $g: \mathcal{U} \times \mathcal{Y} \rightarrow \mathcal{\hat{S}}$ such that $\mathrm{E}[\hat{d}(S, \hat{S}) \leq D]$. 
\end{theorem}

In Wyner-Ziv coding, typical source codewords are split into bins, and only the bin index is forwarded to the decoder. By leveraging the side information, the receiver is able to identify the corresponding compression codeword within the selected bin. Wyner-Ziv coding has been exploited in image and video compression \cite{ASG03, aaron2002wyner}. While in general having the side information available at both the encoder and the decoder is beneficial, for some source-distortion measure pairs, e.g., Gaussian sources under squared-error distortion, it is known that having the side information available only at the decoder does not result in any performance loss \cite{WZ76}. Side information can also be considered for the remote source coding problem since any remote compression problem is equivalent to a standard source coding problem with a new distortion measure as in \eqref{eq:new_distortion}. Side information can also be incorporated into multi-terminal source coding problems, e.g., \cite{GW74}. For example, in the CEO setting in Fig. \ref{f:CEO}, letting one of the observers to have a link to the CEO with sufficiently large rate allows conveying this observation perfectly, which then acts as side information.

\subsection{The Information Bottleneck (IB) and Goal Oriented Compression}\label{ss:IB}

IB  was introduced by Tishby {et    al.}~\cite{TPB99} as a methodology for extracting the information that a variable $S \in \mathcal{S}$ provides about another one $V \in \mathcal{V}$ that is of interest and not directly observable by mapping $S$ into a representation $U\in \mathcal{U}$, as shown in Figure~\ref{f:remoteRD}. Specifically, the IB method consists in finding the mapping $P_{U|S}:\mathcal{S}\rightarrow\mathcal{U}$ that, given $S$, outputs the representation $U$ that is maximally informative about $V$, i.e., such that the mutual information $I(U;V)$  is maximized, while being minimally informative about $S$, i.e., the mutual information $I(U;S)$ is minimized. 
Here, $I(U;V)$ is referred to as the \textit{relevance} of $U$ and $I(U;S)$ is referred to as the \textit{complexity} of $U$, where  complexity is measured by the minimum description length (or bit-length) at which the observation is represented. For a distribution $P_{S,V}$, the optimal mapping of the data with parameter $\beta$, denoted by $P_{U|S}^{*,\beta}$,  is found by solving the IB problem, defined as 
\begin{equation}
L^{\mathrm{IB},*}_{\beta}(P_{S,V}):=\max_{P_{U|S}} I(U;V)-\beta I(U;S)
\label{eq:IBCriteria},
\end{equation}
over all mappings $P_{U|S}$ that satisfy the Markov chain $U - S - V$, where $\beta$ is a positive Lagrange multiplier that allows to trade-off relevance and complexity.
In Section \ref{sec:SolutionsIB} several methods are discussed to obtain solutions to the IB problem in \eqref{eq:IBCriteria} in several scenarios, e.g., when the distribution of  $(S,V)$ is perfectly known or only samples from it are available.   


The IB problem is connected to multiple source-coding problems including source coding with logarithmic loss distortion\cite{CW14}, information combining~\cite{SSZ05,LH06}, common reconstruction\cite{S09}, the Wyner-Ahlswede-Korner problem~\cite{W75b,AK75}, the efficiency of investment information~\cite{EC98}; to communications and cloud radio access networks (CRAN) \cite{E-AZCS19a}, as well as learning, including generalization\cite{VPV18}, variational inference \cite{AFDM17}, representation learning and autoencoders\cite{AFDM17}, neural network compression \cite{dai2018compressing}, and others. See \cite{zaidi2020information} and \cite{GP2020} for recent surveys on the IB principle and its application to learning. The connections between these problems allow extending results from one setup to another, and to consider generalizations of the classical IB problem to other setups including multi-terminal versions of the IB \cite{QPD17,VVP19,  E-AZ19a}. 

In fact, it is now well-known~\cite{HT07} that the IB problem in  \eqref{eq:IBCriteria} is essentially a remote point-to-point lossy source-coding problem~\cite{DT62,WW75,W80}, where the distortion between the desired feature $V^m\in \mathcal{V}^m$ and the reconstruction $\hat{V}^m\in \mathcal{\hat{V}}^m$ is measured under the logarithmic loss (log-loss) fidelity criterion \cite{CW14}. That is,  given $m$ i.i.d. samples $(v^m,s^m)$, an encoder $f^{(m)}:\mathcal{S}^m\rightarrow [1:2^{mR})$ encodes the observation $s^m\in\mathcal{S}^m$ using at most $R$ bits per sample to generate an index $W=f^{(m)}(s^m)$. Using a decoder $g^{(m)}:[1:2^{mR})\rightarrow \mathcal{\hat{V}}$, the receiver generates an estimate $\hat{v}^m\in \mathcal{\hat{V}}^m$ as a probability distribution on $\mathcal{V}^m$  given index $w$ generated from $s^m$. The discrepancy between $v^m$ and the estimate $\hat{v}^m$  is measured using the $m$-letter distortion $d_{\log}(v^m,\hat{v}^m):=\frac{1}{m} \sum_{i=1}^{m} d_{\log}(v_i,\hat{v}_i)$, where
 \begin{equation}
d_{\log}(v,\hat{v}) := \log \frac{1}{\hat{v}(v)},
\label{definition-per-letter-log-loss-distortion-measure}
\end{equation}
where $\hat{v}(v)$ is the value of that distribution evaluated at $v \in V$. 
The decoder $g^{(m)}$ is interested in a reconstruction to within an average distortion $D$, such that
$\mathbb{E}[\frac{1}{m} \sum_{i=1}^{m} d_{\log}(V_i,\hat{V}_i)] \leq D$.
The rate distortion function of this problem can be characterized as follows.

\begin{theorem}(Remote Rate-Distortion Function with Log-Loss Distortion, \cite{DT62,W80})
The rate-distortion function for the remote source coding problem with log-loss distortion measure is given by 
\begin{equation}
    R (D) = \min_{P_{U|S}:D \geq H(V|U)} I(U;S),
\end{equation}
where the optimization is over all distributions satisfying $ V - S - U$.
\end{theorem}

Using the substitution $\Delta := H(V) - D$, the region of achievable pairs $(R,D)$ described by this function can be seen to be equivalent to the convex hull of all pairs $(I(S;U),I(V;U))$ obtained by solving the IB problem in \eqref{eq:IBCriteria} for all $\beta$. Note that, for a given encoder $P_{U|S}$ and reconstruction given by a distribution $Q_{V|U}$ on $\mathcal{V}$, we have
\begin{eqnarray}
\mathbb{E}_{P_{S,V}} [d_{\log}(V,Q_{V|U})] & =& \sum_{s \in \mathcal{S}, \: v \in V} P_{S,V}(s,v) \log\Big(\frac{1}{Q_{v|u}(v|u)}\Big) \nonumber\\
&=& H(V|S) + D_{KL}\big(P_{V|S}\|Q_{V|U}\big), 
\label{optimal-decoder-no-compression}
\end{eqnarray}
which is minimized iff the estimate $Q_{V|U}$ is given by the true conditional posterior $P_{V|U}$. 
Thus, operationally, the IB problem is equivalent to finding an encoder $P_{U|S}$ which maps the observation $S$ to a representation $U$ that satisfies the bit-rate constraint $R$, and such that $U$ captures enough relevance of $V$ so that the posterior probability of $V$ given $U$ minimizes  the KL divergence between $P_{V|S}$ and the estimation $P_{V|U}$ produced by the decoder.


The IB is deeply linked to goal oriented compression, i.e.,  the compression is intended to perform a task. Consider the scenario in which a picture is taken at an edge device and a computational task, such as classifying an element in the picture or retrieving similar images to the one taken, needs to be performed at a remote unit. In classical compression, the image is compressed with the goal of preserving the maximum reconstruction quality before forwarding it over the communication channel. However, in goal-oriented compression only the information or features that are most relevant to perform the task need to be transmitted. For classification, those features should be the most discriminative ones, and not necessarily a representation that allows to reconstruct the image, while for image retrieval, the task is different and so are the most relevant features. That is, the features are specific for the task, and the metric under which the task will be evaluated. This communication scenario is similar to that in Fig.~\ref{f:remoteRD}, where $S$ is a random variable modeling the observation, which is jointly distributed with the relevant information $V$ for the task, which is not directly observable. The goal of the receiver is to recover an estimate of $V$ with sufficient quality to perform the task, while the goal of the transmitter is to encode $S$ into the representation $U$ that conveys only the information necessary to recover $V$ at the receiver, but not necessarily $S$. In the classical setup, in which the image needs to be reconstructed with minimum distortion we have $V=S$, while in classification, $V$ can be the label class to classify the image to, such that the output obtained from the IB is the probability of the observation $S$ belonging to the class given by $V$. 
How to select which are the relevant features for a given task is an open problem and depends on the metric that is used. However, in practise, often a careful encoding of the task into $V$ and a conditional probability tailored for the task estimated by the IB to maximize the relevance can be a good candidate. This is justified, since the log-loss and the mutual information can be used to bound the performance of certain tasks, e.g., the probability of mis-classification of a classifier using a decision rule $Q_{\hat{V}|U}$, denoted by $\epsilon_{V|U}(Q_{\hat{V}|U})$, can be shown to be upper bounded  as $\epsilon_{V|U}(Q_{\hat{V}|U}) := 1- \mathrm{E}_{P_{S,V}}[Q_{\hat{V}|U}]\leq 1-\exp\left( -\mathrm{E}_{P_{S,V}}[d_{\log}(V,Q_{\hat{V}|S})] \right)$ \cite{VPV18}.

\begin{figure*}[t]
\centering
\includegraphics[width=0.85\textwidth]{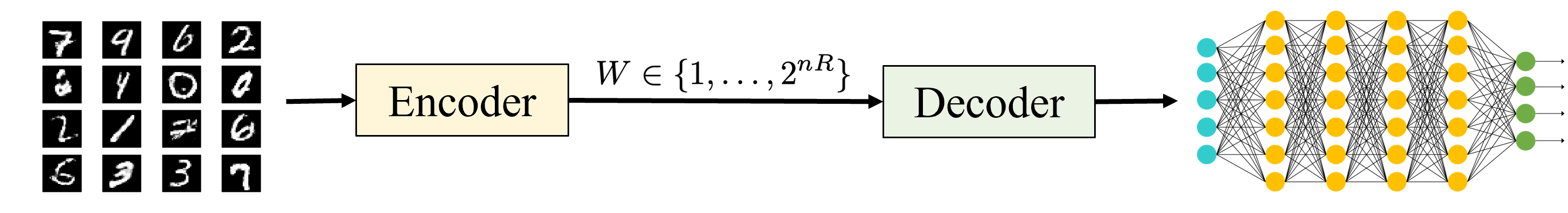}
\caption{Remote model training over a finite-rate communication link.}
\label{fig:remote_training}
\end{figure*}

The formulation of the IB as a remote source coding problem under the log-loss distortion measure can be extended to consider the context in the form of side information at the decoder or both the encoder and decoder \cite{WZ76} as described in Section \ref{ss:goal_oriented} and studied in \cite{UEZ20}. The IB problem can also be extended to multi-terminal scenarios. In particular, in~\cite{E-AZ18a,E-AZ19a}, the distributed  classification problem is studied from an information-theoretic perspective using the IB formulation. In this scenario, one is interested in performing a task at a remote destination, e.g., classification, represented by the latent random variable $V$ using the information relayed by $K$ encoders, each observing some correlated information $(S_1,...,S_K)$ with $V$. As shown in Fig. \ref{f:CEO}, Encoder $k$  encodes the observation $S_k$ into a representation $U_k$ in order to preserve the most relevant and complementary information to the other encoders for the task. This problem can be shown to be essentially a $K$-encoder CEO problem under log-loss distortion \cite{CW14}, in which the decoder is interested in a soft estimate of $V$ from the descriptions $U_1,...,U_K$, each encoded with an average finite rate of $R_k$ bits per sample. The fundamental limits in this scenario can be characterized in terms of the optimal trade-off between relevance $\Delta$ and rate at each encoder as follows.

\begin{theorem}{(Distributed IB Problem\cite{E-AZ19a})}
The relevance-complexity region for the distributed IB  problem is given by the union of all non-negative tuples $(\Delta, R_1,\ldots, R_K)$ that satisfy 
\begin{equation}
\Delta \leq \sum_{k\in \mathcal{S}} [R_k\!-\!I(S_{k};U_{k}|V,T)]  + I(V;S_{\mathcal{S}^c}|T), \quad \forall S \subseteq K
\label{eq:ComplexityrelevanceFunction}
\end{equation}
for some distribution $p_T p_V \prod_{k=1}^K p_{S_k|V} \prod_{k=1}^{K}p_{U|S_k,T}$. 
\end{theorem}

The above IB framework can be used to design encoders and decoders in order to extract the most relevant and complementary information from distinct observations. In Section \ref{sec:SolutionsIB} we describe how to design encoders and decoders by solving (or approximating) the IB problem both for scenarios in which the distribution is perfectly known or when the source distribution is unknown and only data samples are available.


More generally, the approach to extract the most relevant information for a task can also be considered for other end-to-end metrics that go beyond log-loss by modelling the relevant information extraction as a remote source coding problem in which a metric needs to be minimized under some rate constraints, and the metric  is a distortion (not necessarily additive) that represents the performance of the task, and can include, for example, the KL divergence, classification, hypothesis testing, etc. In other cases, the metric can also be learned implicitly by defining an alternative task, as in GANs\cite{NIPS2014_5ca3e9b1}, in which the generator and discriminator are trained to simultaneously extract the relevant information to generate realistic signals by learning how to pass an hypothesis test on distinguishing the generated data from real data.

\subsection{Rate-Limited Remote Inference}\label{ss:remote_inference}

As mentioned above, the hidden feature variable $V$ can represent the class that the sample $S$ belongs to, or the value associated to it in the case of a regression problem. If the goal is to convey only the class information to the receiver, and if there are no constraints on the complexity of the encoder, then the optimal operation would be to detect the class at the encoder, and only convey the class information to the receiver.  From an information theoretic perspective, this problem can also be formulated as a remote inference problem. 

In statistical inference problems, an observer observes $n$ i.i.d. samples $S^n = (S_1, \ldots, S_n)$ from some distribution $p_S$. We assume that this distribution belongs to a known family of distributions indexed by $V \in \mathcal{V}$; that is, $S^n$ follows the conditional distribution $p_{S|V=v}(S^n) = \prod_{i=1}^n p_{S_i|V=v}(s_i)$. In general, the observer may want to estimate $V$ simply from the available samples. If set $\mathcal{V}$ is a discrete set, we have a detection/ \textit{hypothesis testing (HT)} problem. If, instead, $\mathcal{V}$ is a continuous set, i.e., $\mathcal{V} = \mathds{R}^d$, then we have a \textit{parameter estimation} problem. We impose a loss/ distortion function to quantify the quality of the estimation: $l(V, \hat{V})$, where $\hat{V}$ is the estimate of the observer. The expected loss/risk of a decision rule $g^{(n)}: \mathcal{S}^n \rightarrow \mathcal{V}$, where $\hat{V} = g^{(n)}(S^n)$, is then defined as $R_v(g^{(n)}) = \mathds{E}[l(v, \hat{V})]$.

In the Bayesian setting, we assume some known prior distribution on $V$, and try to minimize the average loss (also called \textit{risk}) over the joint distribution of $V$ and $S$, i.e., $\mathds{E}[l(V, g^{(n)}(S^n))]$. Alternatively, we can also aim at minimizing the worst case loss/risk $R_{max} = \sup_{v \in \mathcal{V}} \mathds{E}[l(v, \hat{V})]$. The corresponding decision rule is called the \textit{minimax rule}. The classical Bayesian and minimax inference problems deal with centralized decision problems; that is, they assume the observer and the decision maker are the same agent, and makes the decision with full access to the samples. However, in many practical problems of interest, the observer and the decision maker are connected through a constrained communication channel. If we consider a rate-limited link, we obtain a remote inference problem. Remote inference problems over a rate-limited channel were first considered by Berger in \cite{Berger_1979}. We note that, the remote inference problem in the Bayesian setting is very similar to the information theoretic remote-rate distortion formulation in Section \ref{ss:goal_oriented}, with the exception that we only have a single realization of the latent/hidden variable $V$, rather than a sequence of i.i.d. realizations each generating a separate sample $S_i$. 

When the dimension of the parameter to be estimated is smaller than that of the observations, for example, in the case of HT, the observer can perform local inference and transmit its decision (indeed, optimal performance can be achieved asymptotically at \textit{zero rate} by conveying the \textit{type} information, which is a sufficient statistic \cite{Han-Amari-1989}). Distributed parameter estimation with multiple terminals, each observing a component of a family of correlated samples, was studied in \cite{Zhang-Berger}, under rate constraints from the observers to the decision maker in bits per sample. A single-letter bound, similar to the Shannon's rate distortion function, is provided on the variance of an asymptotically unbiased estimator, which is later improved in \cite{HanAmari:IT:95}. Ahlswede and Burnashev studied the remote estimation problem when the decision maker has its own side information \cite{Ahlswede-Burnashev}. 

Ahlswede and Csisz\'{a}r studied HT when the decision maker also has its own observations \cite{Ahlswede-Csiszar}. They studied the exponent of the type-II error probability, when a constrained is imposed on the type-I error probability. For the case of testing against independence, they were able to provide a single-letter expression similar to Shannon's rate-distortion function. This is one of the few cases in which a single-letter characterization is possible for a non-additive distortion measure. The more general distributed setting is considered in \cite{Han:TIT:87}. Han showed in this paper that a positive exponent can be achieved even with a single-bit compression scheme. This result was extended to the more general zero-rate compression in \cite{Han:TIT:89}. This result was later refined by Shalaby and Papamarcou in \cite{Shalaby:TIT:92}, where they show that when the observers have fixed codebook sizes, the asymptotic performance does not depend on the particular codebook size. This means that no further gain can be obtained in terms of the asymptotic error exponents by allowing each observer to transmit a high-resolution soft decision instead of a binary decision. Despite significant research efforts, the optimal characterization of the type-II error exponent for the remote HT problem for the general case (beyond testing against independence) remains open to this day. Lower bounds are provided for the general problem in \cite{Ahlswede-Csiszar} and \cite{Han:TIT:87}. Distributed hypothesis testing is also studied in the context of a sensor network in \cite{Chamberland:TSP:03}, where multiple sensors convey their noisy observations to a fusion center over rate-limited links. There has been a recent resurgence of interest in distributed HT problems \cite{Rahman:TIT:12, Katz-Pablo, Sadaf-Wigger-Li, WK-2017, Escamilla:TIT:20}.

\section{Machine Learning Techniques for Semantic- and Task-Oriented Compression}\label{s:ML_Semantics}

The ultimate motivation of semantic compression is to extract the semantic information within the source data at the transmitter that is most relevant for the task to be executed at the receiver. By filtering out task-irrelevant data both the bandwidth consumption and the transmission latency can be reduced significantly. However, the information theoretic framework presented above either assumes known statistics for the data and the relevant features for the task, or it is limited to the parameter estimation framework assuming i.i.d. samples from a family of distributions. On the other hand, in most practical applications we do not have access to statistical information, and often need to make inferences based on a single data sample. An alternative approach is to consider a data-driven framework, where we have access to a large dataset, which would allow us to train a model using machine learning tools to facilitate semantic information extraction without requiring a mathematical model. In particular, deep learning (DL) aided semantic extraction techniques have shown great potential for various information sources and associated tasks in the recent years. 

While most machine learning research can be considered within the context of semantic feature extraction, we will focus on the communication aspects here. Machine learning algorithms typically follow a two-step approach: in the training phase, a model is trained using the available dataset for the desired task, e.g., classification or regression. Once the model is trained, it is used for prediction on new data samples. Communication in both phases can be considered in the context of semantic or goal-oriented communications. Below, we overview research in these two phases separately. 

\subsection{Remote Model Training}\label{ss:Compress4Training}

In the training phase, a single node or multiple nodes each with its own dataset communicate with a destination node with the goal to reconstruct a model at the destination for a particular inference task. Note that this may also correspond to a storage problem, where the goal is to store the model in a limited memory to be later used in predicting future data samples. We would like to highlight here that this problem is an instance of a particular remote rate-distortion problem. Let us consider first the point-to-point version of the problem illustrated in Fig. \ref{fig:remote_training}. Here, we can treat the dataset at the encoder as the information available at the encoder, and the model itself as the remote source that the decoder wants to recover. Note that, similarly to the other semantic rate-distortion problems, the fidelity measure here is also not the similarity of the reconstructed neural network weights at the receiver to those trained at the encoder. In the end, what really matters is the performance of the reconstructed model at the decoder in terms of the prescribed quality measure, e.g., the accuracy of the reconstructed model at the decoder on new data samples. 

As in the previous remote rate-distortion problems, a natural solution approach is to first estimate the remote source at the encoder; that is, to first train a model locally, and then convey this model to the decoder with the highest quality over the rate-limited channel, that is, in a way retaining the predictive power of the model on future queries. While the former step is simply the standard training process, the latter corresponds to model compression, which has been widely studied in recent years particularly in the context of deep neural networks (DNNs) that would otherwise require significant communication or storage resources. 

\begin{figure}[t]
\centering
\includegraphics[width=0.43\textwidth]{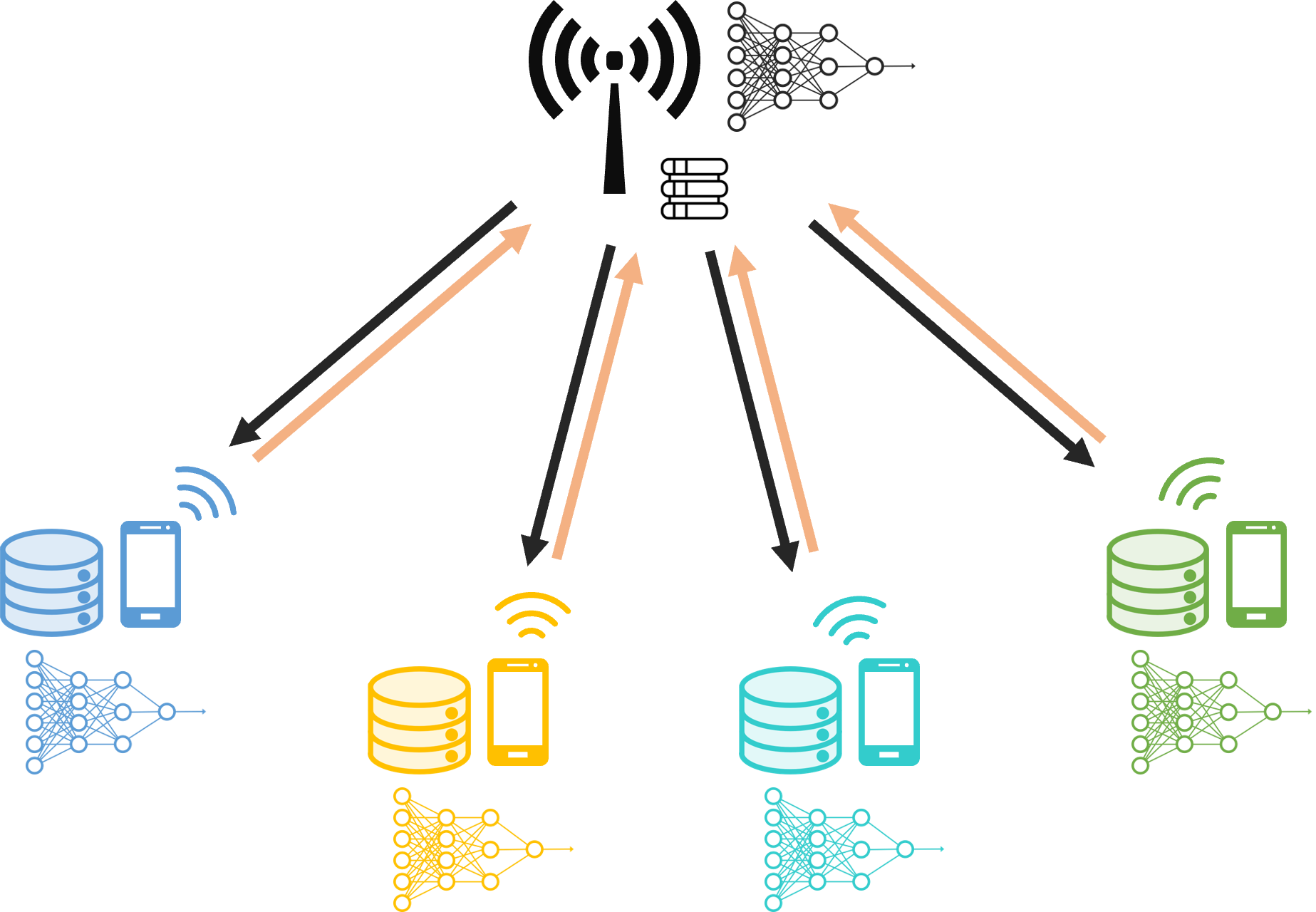}
\caption{Federated model training.}
\label{fig:federated_training}
\end{figure}

There are various widely used methods to reduce the size of a pre-trained model. These include parameter pruning,  quantization, low-rank factorization \cite{Denton14}, and knowledge distillation. It has been known for a long time that many parameters in a neural network are redundant, and do not necessarily contribute significantly to the performance of the network. Therefore, redundant parameters that do not have a significant impact on the performance can be removed to reduce the network size and help address the overfitting problem \cite{Hanson:NIPS:89, LeCun:NIPS:90, Hassibi:NIPS:93}. In particular, the so-called `optimal brain surgeon' in \cite{Hassibi:NIPS:93} uses the second-order derivative, i.e., the Hessian, of the loss function with respect to the network weights. In \cite{LeCun:NIPS:90}, the authors assumed that the Hessian matrix is diagonal, which causes the removal of incorrect weights. In \cite{Hassibi:NIPS:93}, the authors showed that the Hessian matrix is non-diagonal in most cases, and they proposed a more effective weight removal strategy. In addition to its better performance, the optimal brain surgeon does not require re-training after the pruning process. Note that pruning the network would also reduce the complexity and delay of the inference phase. Pruning is still a very active research area, and many different pruning methods have been studied, including weight, neuron, filter, or layer pruning. Please see \cite{liu2020pruning} and references therein for a detailed survey on advanced pruning techniques. Another approach is to train the network directly with the sparsity constraints imposed during training, rather than reducing the network to a sparse one after training a full network  \cite{Lebedev_2016_CVPR, Wen:NeurIPS:16}. 

We can also apply quantization or other more advanced compression techniques, e.g., vector quantization, on the network parameters. Quantization has long been employed to reduce the network size for efficient storage \cite{Fiesler90, BALZER91}. It is well-known that low precision representation of network weights is sufficient not just for inference based on trained networks but also for training them \cite{Gupta:JMLR:15, courbariaux2014training}. At the extreme, it is possible to train DNNs even with single-bit binary weights \cite{Courbariaux:NIPS:15, Courbariaux:arXiv:16}. 

DNN compression can also be treated as a standard source compression problem, and vector quantization techniques can be employed for codebook-based compression to reduce the memory requirements. Similarly to pruning, Hessian-based quantization is shown to be effective in \cite{Choi:ICLR:17}. Hash functions are employed in \cite{Chen:ICML:15}, where the connections are hashed into groups, such that the ones in the same hash group share weights. It is argued in \cite{Gong:arXiv:14} that, for a typical network about 90\% of the storage is taken up by the dense connected layers, while more than 90\% of the running time is taken by the convolutional layers. Therefore, the authors focus on the compression of dense connected layers to reduce the storage and communication resources, and employ vector quantization to reduce the communication rate. In \cite{han2016deep}, Huffman coding is used to further compress the quantized network weights.

In knowledge transfer, the goal is to transfer the knowledge learned by a large, complex ensemble model into a smaller model without substantially reducing the network performance. It was first studied in \cite{Bucilau:06}. In \cite{hinton2015distilling}, the concept of temperature was introduced to generate the soft targets used for training the smaller model.

Another possible approach to solve this problem is model architecture optimization, where the goal is to adjust the size and complexity of DNN architectures to the constraints of the communication link during training without sacrificing their performance. Some of the popular recent efficient model architectures include SqueezeNet \cite{Iandola:SqueezeNet}, MobileNets \cite{howard2017mobilenets}, ShuffleNet \cite{Zhang:ShuffleNet}, and DenseNet \cite{Huang:DenseNet}. We refer the reader to \cite{Deng:Survey} for a more comprehensive survey of recent advances in model compression techniques for DNNs.

A more common scenario in remote model training is distributed training, in which multiple nodes each with its own local dataset collaborate to train a model by communicating with a remote parameter server, or with each other. The former scenario is known as \textit{federated learning}, while the latter is referred to as \textit{fully distributed}, or \textit{peer-to-peer learning}. Please see Fig. \ref{fig:federated_training} for an illustration of the federated learning scenario. Similarly to the single node scenario discussed above, federated training can be treated as a multi-terminal rate-distortion problem, where the datasets are observed samples at the multiple encoders, correlated with the underlying model, which is to be recovered at the parameter server. This would correspond to the CEO problem presented in Section \ref{ss:goal_oriented}, implemented with multiple rounds of two-way communication between the nodes and the parameter server. Stochastic gradient descent based iterative algorithms are often used for federated learning. In the federated averaging (FedAvg) algorithm, proposed in \cite{McMahan17}, a global model is sent from the parameter server to the nodes, each of which computes a model update, typically employing multiple stochastic gradient descent updates. The nodes then transmit these model updates back to the parameter server, which aggregates them, to finally update the global model. The algorithm is iterated until convergence. At each iteration of the algorithm, the goal is then to compute the average of the model updates, rather than the individual updates. Hence, this is a distributed lossy computation problem, which can be considered as yet another aspect of semantic communications. Here, the semantic that is relevant for the underlying task is a function of the multiple signals observed at different nodes. 


Computation is often considered as a distinct problem from communication. One approach to computation over networks would be to carry out separate communication and computation steps. For example, if a node wants to compute a function of random variables $S_1, \ldots, S_n$ that are distributed over the network, we can first deliver these random variables to the node, which then computes the function value. In the point-to-point setting, the optimality of this approach can be shown in certain cases following the arguments of remote rate-distortion problem, where we treat the function to be computed as the latent variable $V$ of the observed source $S$ (see Fig. \ref{f:remoteRD}). 
However, in the general case, the optimal performance for a generic function is an open problem, even in the lossless computation case. The multi-terminal function computation problem was first introduced in \cite{Korner1979}, where the authors considered the parity function of two correlated symmetric binary random variables. They identified the optimal rate region for this case, and showed that this is not equivalent to the rate region one would obtain from \cite{B77} by first compressing and sending the observed sequences to the decoder. This illustrates the difficulty of the problem for arbitrary function computation. The problem was later studied in \cite{Orlitsky:TIT:01} in a point-to-point setting, considering one of the two sources is available at the decoder as side information. The optimal rate required for lossless computation of any function (in the Shannon theoretic sense - over long blocks with vanishing error probability) is characterized, and is shown to be given by the the conditional $G$- entropy \cite{simonyi1995graph} of $X$ given $Y$, where $G$ is the characteristic graph of $X, Y$ and function $g$ to be computed as defined in \cite{Witsenhausen:TIT:76}. While this is in general lower than first sending $X$ to the decoder at rate $H(X|Y)$, and then computing the function, it is observed in  \cite{Orlitsky:TIT:01} the gain is marginal in most cases. 

\subsection{Remote Inference}\label{ss:remote_inference_ML}

We next consider the machine learning approaches for rate-limited remote inference problems. Following the arguments in Section \ref{s:semantic_compression}, various lossy compression problems can be considered in the context of semantic communication under the appropriate reconstruction metric. In recent years, DNN aided compression algorithms have achieved significant results, often outperforming state-of-the-art standardised codecs in a variety of source domains, from image  \cite{TodericiICLR2016, RippelICML2017, Balle:ICLR:17, Balle:NIPS2018, minnen2020channelwise}, to video \cite{agustsson_scale-space_2020,wu_video_2018,lu_dvc:_2018,rippel_learned_2018,djelouah_neural_2019,habibian_video_2019,han_deep_2019,hu_fvc_2021,liu_deep_2021}, speech \cite{Kankanahalli:ICASSP:18, Kleijn:ICASSP:18}, and audio \cite{Petermann2021HarpNetHR, Zeghidour:TASLP:22} compression. One of the main advantages of DNN-based approaches compared to conventional compression algorithms is that they can be trained for any desired reconstruction metric at the receiver. For example, image or video compression algorithms can be trained with the SSIM or MS-SSIM metrics as objective, providing perceptually better reconstructions. We refer the reader to \cite{yang:neural_compression} for a comprehensive overview of recent developments in both lossless and lossy compression using DNNs and other machine learning methods. 

Task-oriented image compression was considered in \cite{1996SPIE:Anderson}, where the authors proposed lossy compression of MRI images to preserve as much clinically useful information as possible depending on the diagnostic task to be performed. In \cite{Pu:ICIP:14}, the authors propose a metric based on conditional cross entropy for a target detection task. In video compression, one approach is to employ region-of-interest compression, where only the relevant part of the video stream is compressed \cite{Bulla2013RegionOI, Xu:STSP:14, Andalo:16}. A classification aware distortion metric is proposed in \cite{Minervini:VCIP:15}, and applied to the high efficiency video coding (HEVC) standard.

The authors of \cite{torfason2018towards} have shown that latent representation produced by compressive autoencoders can be used to perform a classification task with ResNet-$50$ \cite{resnet} network resulting in almost the same accuracy obtained by training on uncompressed image, showing that the classification network does not need to reconstruct the image first, at least explicitly. They also consider jointly training of the compression and classification networks. Task-based quantization is studied in \cite{Shlezinger2021DeepTQ} in the context of analog-to-digital conversion of signals for a specific task, and for channel estimation in \cite{Shohat:ICASSP:19}.

\begin{figure}[t]
\centering
\includegraphics[width=0.43\textwidth]{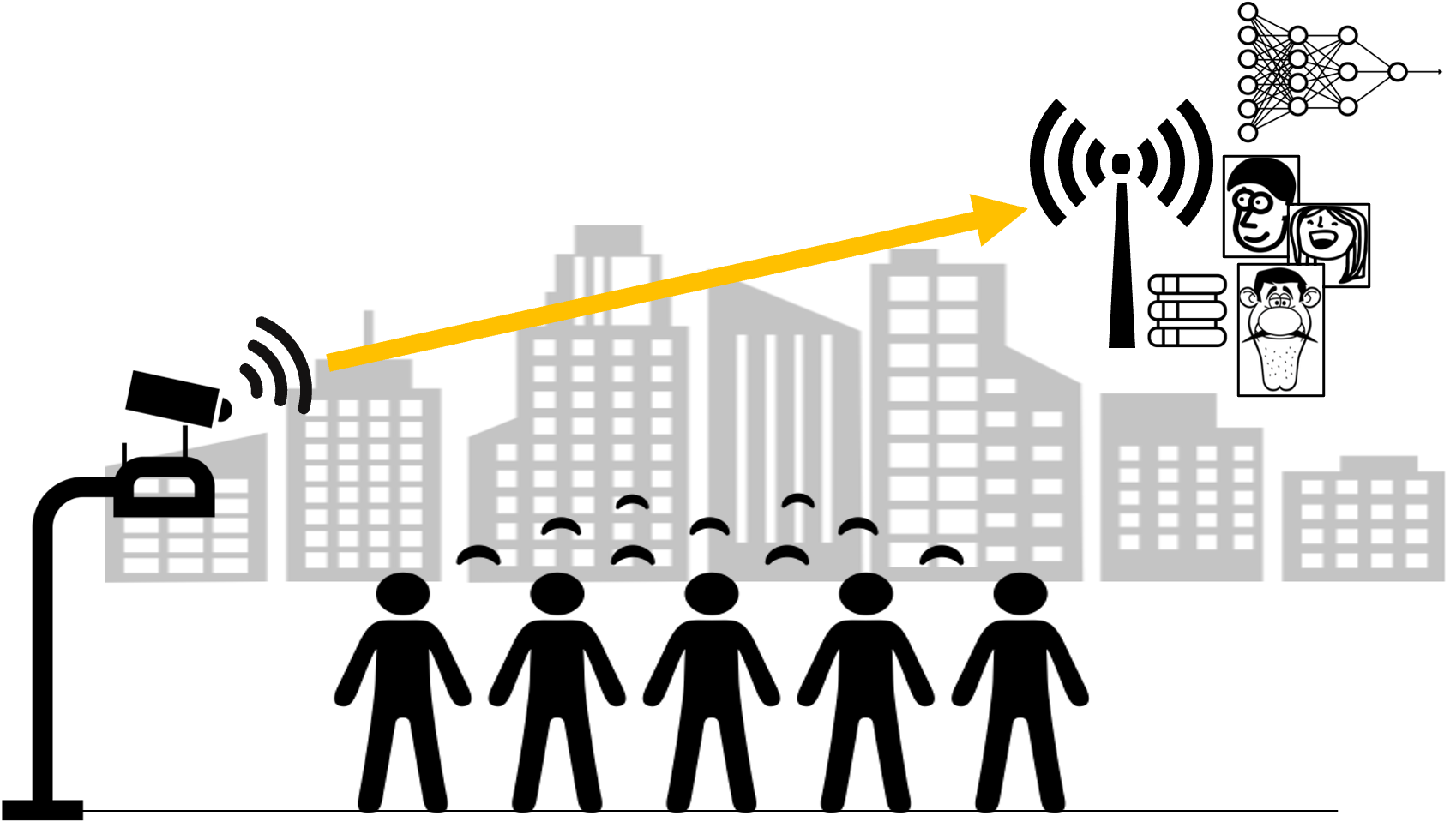}
\caption{Remote image retrieval problem.}
\label{fig:retrieval}
\end{figure}

We remark that the aforementioned task-based compression problems are remote rate-distortion problems in essence, but machine learning tools are employed mainly to acquire statistical knowledge from data. However, in these cases, since the transmitter has access to the original source information, the desired task can often be carried out at the transmitter, which is then transmitted to the receiver over the channel. Particularly, for classification tasks this would require only a limited amount of information to be transmitted. \highlight{However, complete classification at the transmitter may not be possible due to complexity constraints, e.g., when the transmitter is a simple IoT device. In such a scenario, the transmitter may extract some features, which are then conveyed to the decoder using a finite number of bits, and the rest of the inference task is carried out at the receiver end. This is called as `split learning' in the literature. In the context of DNNs, split learning refers to dividing a DNN into two parts, the head and the tail. The head consists of the first layers of the DNN architecture that are executed at the encoder, while the tail consists of the later layers that are executed at the receiver. From a rate-distortion perspective, the goal is to convey the features obtained at the end of the head network to the receiver with as few bits as possible, while still achieving the desired inference accuracy. Quantization and/or compression of feature vectors is considered in \cite{Li:ICANN:18, JALAD:18} and \cite{BottleNet:19}. While the former considered using a quantized version of the network at the transmitter side to also reduce the storage and computation requirements, the latter considered split learning also for unsupervised learning with an autoencoder architecture. Ideas from knowledge distillation and neural image compression are exploited in \cite{Matsubara:WACV:22} to obtain a more efficient compression scheme for the intermediate feature representations obtained by the head network.}

A different type of remote inference problem is considered in \cite{Jankowski:JSAC:21}, called image retrieval at the edge. In this setting, illustrated in Fig. \ref{fig:retrieval}, the goal is to identify a query image of a person or a vehicle recorded locally by matching with images stored in a large database (gallery), typically available only at the edge server. We emphasize that, unlike the typical classification tasks, the retrieval task cannot be performed locally as the database is available only at the remote edge server. In \cite{Jankowski:JSAC:21}, the authors propose a retrieval-oriented image compression scheme, which compresses the feature vectors most relevant for the retrieval task, depending on the available bit budget. To reduce the communication rate, the authors quantize and entropy code the features to be transmitted, using a learned probability model for the quantized bits for efficient compression.

\highlight{In \cite{NEURIPS2019_6c29793a}, the classification-distortion-perception trade-off is studied, assuming that the reconstructed image at the receiver is also fed into a prescribed classification network. It is shown that the classification error rate on the reconstructed signal evaluated by the prescribed classifier cannot be made minimal along with the distortion and perceptual measures. A similar semantic-oriented compression approach is applied to facial image compression in \cite{Chen:N:19}, where regionally adaptive pooling is used to optimize the compression parameters according to gradient feedback from the hybrid distortion-perception-semantic fidelity metric. It is shown that the semantic distortion metric allows allocating more bits for the compression of more semantically critical areas in face images. Automatic generation of semantic importance maps is considered in \cite{Li:TIP:21}, where the output of instance segmentation (combination of object detection and semantic segmentation) through Mask-RCNN \cite{He:MaskRCNN} is used as the importance measure of each segment, and the bit allocation is carried out using reinforcement learning.}



\section{Semantic- and Task-Oriented Communication over Noisy Channels: A JSCC Approach}\label{s:JSCC}

In the previous section, we have mainly focused on the compression aspects, assuming an error-free finite-rate communication channel from the encoder to the decoder. However, many communication  channels suffer from noise, interference, and other imperfections. Shannon's channel coding theory mainly deals with communication over such noisy communication channels. However, as we have mentioned earlier, channel coding theory focuses on the reliable delivery of bits, whereas in the context of semantic communication, we will consider the transmission of source signals such as image, video, audio, or their features relevant for a particular task, over a noisy channel.

\begin{figure}
		\centering
		\includegraphics[width=3.5in]{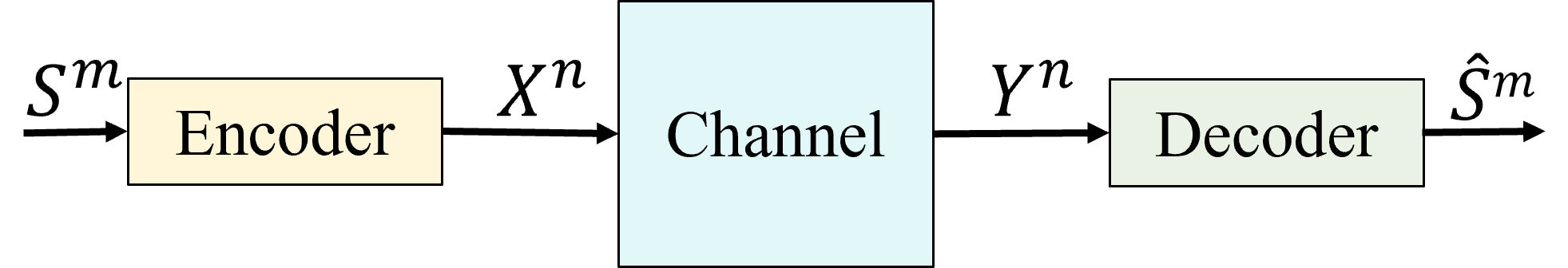}
		\label{f:JSCC_model}
		\caption{Illustration of a JSCC problem over a noisy communication channel. }
	\end{figure}

In this problem, illustrated in Figure \ref{f:JSCC_model}, the transmitter wants to transmit a sequence of independent source symbols $S^m \in \mathcal{S}^m$ each sampled from the distribution $p_S(s)$, over a memoryless noisy communication channel characterized by the conditional probability distribution $P(Y|X)$, where $X \in \mathcal{X}$, $Y \in \mathcal{Y}$. Let $\hat{S}^m \in \mathcal{\hat{S}}^m$ denote the reconstruction at the receiver based on $Y^n$. Similarly to the rate-distortion theory formulation, the goal is to minimize the distortion between $S^m$ and $\hat{S}^m$ under some given distortion (fidelity) measure, $d: \mathcal{S}^m \times \mathcal{\hat{S}}^m \rightarrow [0, \infty)$. More formally, let $f^{m,n}:  \mathcal{S}^m \rightarrow \mathcal{X}^n$ denote the encoding function, and $g^{m,n}:  \mathcal{Y}^n \rightarrow \mathcal{\hat{S}}^m$ denote the decoding function. In the case of an average distortion criteria, the goal is to identify the encoder and decoder function pairs that minimize $\mathds{E}[d(S^m, \hat{S}^m)]$, where the expectation is over the source and channel distributions as well as any randomness the encoding and decoding functions may introduce.  One can also impose an excess distortion criterion, where the goal is to minimize $\mathds{P}[d(S^m, \hat{S}^m) >d]$, for some maximum allowable distortion target $d>0$. 

A $(m,n)$ joint source-channel code of rate $r=m/n$ consists of an encoder-decoder pair, where the encoder $f^{(m,n)}: \mathcal{S}^m \rightarrow \mathcal{X}^n$ maps each source sequence $s^m$ to a channel input sequence $x^n(s^m)$, and the decoder $g^{(m,n)}: \mathcal{Y}^n \rightarrow \mathcal{\hat{S}}^m$ maps the channel output $y^n$ to an estimated source sequence $\hat{s}^m$. A rate-distortion pair $(r,D)$ is said to be \textit{achievable} if there exists a sequence of $(m,n(m))$ joint source-channel codes with rate $r$ such that $n(m) \leq r m$, $\forall m$, and 
\begin{equation}
	\limsup_{m\rightarrow \infty} \mathrm{E}[d(S^m, \hat{S}^m(Y^{n(m)}))] \leq D.
\end{equation}

Shannon proved his well-known Separation Theorem for a single-letter additive distortion measure; that is, $d(S^m, \hat{S}^m) = 1/m \sum_{i=1}^m d(S_i, \hat{S}_i)$ for the distortion measure $d(S, \hat{S}) \in [0, \infty)$. The theorem states the following.

\begin{theorem}(Shannon's Separation Theorem, \cite{Shannon}) Given a memoryless source $S$ and a memoryless channel $p_{Y|X}$ with capacity $C = \sup_{p_X(x)} I(X;Y)$, a rate-distortion pair $(r,D)$ is achievable if $r R(D) < C$. Conversely, if a rate-distortion pair $(r,D)$ is achievable, then $r R(D) \leq C$.
\end{theorem}

The theorem states that we can separate the design of the communication system into two sub-problems without loss of optimality, the first focusing on the compression and the second focusing on the channel coding, each of them designed independently of the other. However, the optimality of separation holds only in the limit of infinite blocklength; whereas, in practice, it is possible to design joint source-channel codes that would outperform the best achievable separate code design. This was observed by Shannon in his 1959 paper \cite{Shannon_RD}. Considering a binary source generating independent and equiprobable symbols and a memoryless binary symmetric channel, Shannon observed that simple uncoded transmission of symbols achieves the optimal distortion with rate $r=1$ for one particular value of distortion $D$ determined by the error probability of the channel. This observation was later extended by Goblick in \cite{Goblick:TIT:65} to Gaussian sources transmitted over Gaussian channels. This happens when the source distribution matches the optimal capacity achieving input distribution of the channel, and the channel at hand matches the optimal test channel achieving the optimal rate-distortion function of the source. Necessary and sufficient matching conditions are given in \cite{Gastpar:TIT:03} for general source and channel distributions. However, these conditions are not satisfied for most practical source and channel distributions, and even when they hold, optimality of uncoded transmission fails when the coding rate is not $1$, i.e., in the case of bandwidth compression or expansion. On the other hand, the presence of such optimality results, that is, the fact that asymptotically optimal performance that requires infinite blocklength source and channel codes can be achieved by simple zero-delay uncoded transmission implies that there can be other non-separate coding schemes that can achieve near optimal performance, or outperform separation-based schemes in the finite blocklength regime.

\subsection{Multi-terminal JSCC} 
It is well-known that the optimality of separation does not directly generalize to multi-terminal scenarios, even in the infinite blocklength regime. This observation is often referred to the seminal work by Cover, El Gamal and Salehi \cite{Cover:TIT:80}, where they consider the transmission of correlated sources over a multiple access channel (MAC), and provide an example in which the uncoded transmission of the sources allow their perfect recovery, while this cannot be achieved by a separate scheme. Interestingly, it is much less known that a similar observation was already made by Shannon in \cite{Shannon1961TwowayCC}, where he considers the transmission of correlated sources over a two-way channel. 

The authors of \cite{Cover:TIT:80} also proposed a coding technique (achievability result) exploiting the correlation among the sources. This result showed that instead of removing the correlation, we can utilize the dependency among the sources to design correlated channel codes, and in certain cases transmit the sources reliably even though this would not be possible with distributed compression followed by independent channel coding. Shortly afterwards a counter example was given in \cite{Dueck:TIT:81} showing that the sufficiency conditions provided in \cite{Cover:TIT:80} are not necessary. More recently, \cite{Kang:CISS:06} gave finite-letter sufficiency conditions for the lossless delivery of correlated sources over a MAC. A similar problem of broadcasting correlated sources to multiple users is considered in \cite{Han:TIT:87}, while \cite{ Nayak:TIT:10} considers broadcasting to multiple receivers each with a different distortion measure and side information. Many JSCC transmission strategies have been extensively studied for Gaussian sources over multi-terminal  \cite{Bross2010, Lapidoth2010, Lapidoth2010a}, as well as non-ergodic scenarios \cite{IG16Exponent}.

\subsection{Remote Inference over Noisy Channels}

In Subsection \ref{ss:remote_inference}, we have presented various statistical inference problems under rate constraints. We highlighted that these inference problems do not satisfy the additive single-letter requirement of typical Shannon theoretic distortion measures considered in the context of rate-distortion theory. Therefore, the separation theorem does not directly apply for these distortion measures. 

Distributed hypothesis testing problem over a noisy communication channel was studied in \cite{Sreekumar-Gunduz} considering the type II error exponent (under a prescribed constraint on the type I probability of error) as the performance measure. Here, the task is to make a decision on the joint distribution of the samples observed by a remote observer and those observed by the decision maker. The observer communicates to the decision maker over a noisy channel.  A separate hypothesis testing and channel coding scheme is presented, combining the Shimokawa-Han-Amari scheme \cite{Shimokawa} with a channel code that achieves the expurgated exponent with the best error-exponent for a single special message \cite{Borade-09}. A joint scheme is also proposed using \textit{hybrid coding} \cite{Minero}. It is shown that separate scheme achieves the optimal type II error exponent when testing against independence. This is a special case of the problem, when the  alternate hypothesis is the independence of the samples observed by the observer and the decision maker. 
While the optimal type II error exponent remains open in general, it is shown in \cite{Sreekumar-Gunduz} that joint encoding can strictly improve upon separation. This shows that communication and inference cannot be separated (without loss of optimality), but how the two should interact is vastly unexplored.

Distributed hypothesis testing over independent additive white Gaussian noise (AWGN) and fading channels, respectively, is studied in \cite{Chamberland:JSAC:04} and \cite{Niu:TSP:06}. These papers consider multiple sensors making noisy observations of the underlying hypothesis, and communicate to a fusion center over orthogonal noisy channels. Hypothesis testing over a discrete MAC is studied in \cite{Duman:98}, where the observations are quantized before being transmitted. Distributed estimation over a MAC is studied in \cite{Mergen:TSP:06}, and a type-based uncoded transmission scheme is shown to be asymptotically optimal. Distributed estimation over a MAC is studied in \cite{Lee:JSAC:22} from a worst case risk point of view. Analog/uncoded transmission is again shown to outperform its digital separation-based counterparts, and to achieve a worst case risk that is within a logarithmic factor of an information theoretic lower bound.

\section{Practical Designs for Goal-Oriented Communication over Noisy Channels}

Practical designs for JSCC of various information sources have been a long standing research challenge. Many different designs have been proposed in the literature, mainly based on the joint optimization of the parameters of an inherently separate design \cite{Modestino:TCOM:79, Moore:TCOM:84, Farvardin:IT:90, Farvardin:IT:91, Skoglund:IT:99a, Kozintsev:TSP:98, Skoglund:IT:99b, Cheung:TIP:00, Hochwald:TIT:97}. Another group of JSCC schemes instead consider a truly joint design. Motivated by the theoretical optimality of uncoded transmission in certain ideal scenarios, analog transmission of discrete cosine transform (DCT) coefficients is proposed in \cite{Jakubczak:10} for wireless image transmission. In \cite{Bursalioglu:ITA:11}, the authors proposed linear coding of quantized wavelet coefficients. However, these efforts in JSCC design either do not provide sufficient performance gains, or they are too complex and specific to the underlying source and channel distributions to be applied in practice.

Recently, JSCC schemes based on autoencoders \cite{rumelhart1985learning}, which are DNNs aimed at unsupervised dimensionality reduction for high-dimensional data, have been introduced \cite{Eirina:TCCN:19, farsad2018deep, kurka2019deepjsccf, Kurka:TWC:21, Tung2021DeepWiVeDW}, and are shown to provide comparable or better performance than state-of-the-art separation-based digital schemes. \highlight{Compared to the semantic data compression schemes we have seen in Section \ref{s:ML_Semantics}, the main challenge in semantic communications over wireless channels is the stochasticity in the channel. Therefore, the designed code should not only compress the input signal to the available channel bandwidth, but also design a mapping that can mitigate the effects of channel uncertainties; that is, it should act both as a compression and an error correction code. As we will later highlight, learned end-to-end JSCC schemes for semantic communications (particularly for high-dimensional inputs such as image and video) generate channel inputs that are correlated with the source signal. In a sense, these schemes transform the manifold representing the source signal to another manifold in the channel input space in a continuous manner; that is, similar source signals are mapped to similar channel inputs so that the reconstruction is not far from the input signal even after some noise is added by the channel. Thanks to this correlation, in addition to improving the performance for a fixed channel state, these JSCC schemes also achieve graceful degradation with channel quality. That is, unlike separation-based approaches, their performance does not fall apart when the channel quality falls below a certain threshold (below which the channel code cannot be decoded reliably), but instead gracefully degrades as the channel quality worsens. }

\highlight{We reemphasize that both the source and channel statistics need to be taken into account when designing semantic communication schemes. When training the semantic communication system in an end-to-end manner, the stochasticity due to the wireless channels will affect  the forward-propagation and the back-propagation in the training process. Moreover, in addition to a dataset representing the potential source signals, we also need to be able to model the channel statistics during training. In the case of a finite input channel with a discrete input alphabet, this may be acquired in the form of a dataset of transmitted and received signal pairs over the particular channel, or, in general, we can simply use the model of the channel if an accurate model is available. Another possibility would be to embed the physical channel into the training process, but this will slow down the training significantly. Another consequence of the randomness in the channel is that the reconstruction at the receiver becomes random, as it depends on the channel realization. This is different from conventional digital communication systems, where the source signal is reconstructed as a prescribed representation determined by the deterministic compression scheme as long as the channel codeword is decoded successfully. The channel uncertainty also creates further challenges regarding the availability of channel state information (CSI) and how this information can be used within the learning framework. }

\begin{figure*}[tbp]
\includegraphics[width=0.6\textwidth]{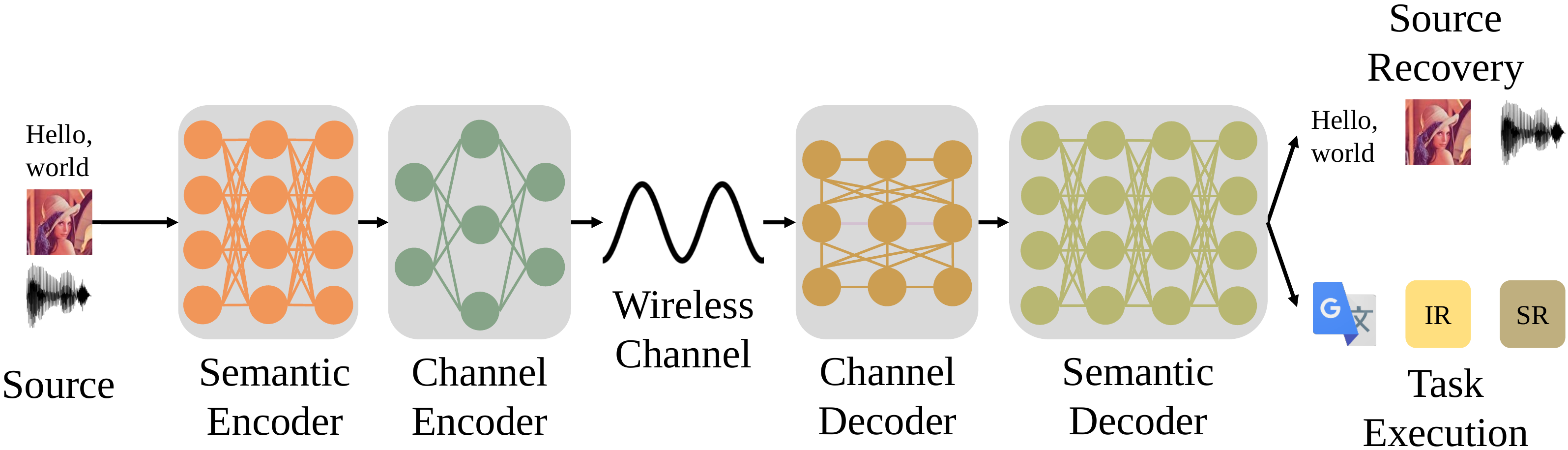} 
\centering 
\caption{Schematic of a DL-enabled semantic communication system.}
\label{general_semantic_communication}  
\end{figure*}

The general schematic of a semantic communication system is shown in Fig.~\ref{general_semantic_communication}. The input data is sequentially passed through a semantic encoder and joint source-channel encoder to extract semantic information relevant to the receiver's task, which can  be either source signal recovery or intelligent task execution. The benefits of this semantic communication approach is due to both the intelligent semantic encoding step, which basically extracts task-relevant features from the input, and the JSCC approach to delivering these features to the receiver. \highlight{On the other hand, depending on the system design approach, the complexity of the data and the downstream task, the distinction between the semantic encoder/decoder and the JSCC encoder/decoder may not be clearly demarcated, and both of them can be implemented through a single DNN architecture, and trained jointly in an end-to-end fashion.}

\highlight{The intelligent task-oriented semantic communications have attracted intensive investigation in recent years due to the capability to address pertinent challenges in the traditional communication system, which has been considered as one of the key technologies to cater to the unprecedented demands of intelligent tasks in the future intelligent communications era. While the JSCC approach can potentially outperform separate source and channel coding, particularly in the short blocklength regime, it loses the modularity. Modularity refers to the separate design of source and channel coding schemes, where the source encoder can be designed oblivious to the channel statistics or the particular channel coding and modulation scheme employed for communication. All the source encoder needs to know is the level of compression, in terms of the bit per source sample. Similarly, the channel code can be designed oblivious to the particular source signal and its statistics. However, in the case of JSCC, since the code is designed in an end-to-end fashion, we need to take the source and channel statistics into account in a joint manner. Therefore, we will introduce task-oriented semantic communications separately according to the different types of source data, for text, speech, and image, respectively.}

\subsection{{Task-Oriented Semantic Communications for Multimodal Data}}
\subsubsection{Task-Oriented Semantic Communications for Text}

For networks that allow interactions between humans as well as smart devices that have unique backgrounds and behavior patterns, reliable communication can be redefined as the {\it intended meaning} of messages being preserved at reception. Further, communicating parties can form social relationships and build trust, which may further affect how the received messages are interpreted. Motivated by these factors, reference \cite{Guler-Yener-percom14} has proposed an approach that takes into account the meanings of the communicated messages and demonstrated the design of a point-to-point link to reliably communicate the meanings of messages through a noisy channel. To do so, in \cite{Guler-Yener-percom14} the authors proposed a novel performance metric, {\it semantic error,} to measure how accurately the meanings of messages are recovered, and then determined the optimal transmission policies to best preserve the meanings of recovered messages. This is achieved by leveraging lexical taxonomies and contextual information to design a graph-based index assignment scheme for fixed rate codeword assignment in a noisy channel. Words that are similar in their meaning (measured by the semantic distance between the words) are assigned to closer codewords in terms of their Hamming distance. 

Building on this work, \cite{Guler-Yener-Swami-tccn} modeled an external agent who can influence how the destination perceives the meaning of received information, to study the impact of social influence on contextual message interpretations on semantic communication. The exact nature of the agent, whether adversarial or helpful, is unknown to the communicating parties. This problem is first modeled as a Bayesian game played between the encoder/decoder and the influential entity in \cite{Guler-Yener-Swami-tccn}. By extending the Bayesian game into a dynamic setting, the authors studied the interplay between the influential entity and the communicating parties, in which each player learned the true nature of the other player by updating its own beliefs as the game progresses, revealing information through observed actions. While these works pre-date the recent efforts that brought semantics into the center stage, they serve as early works to build on designing semantic communication networks.


However, the semantic error of the aforementioned text-based semantic communication system is measured only at the word level. In an early effort on JSCC for text transmission using deep learning techniques, the authors of \cite{farsad2018deep} considered sentence level similarity using the edit distance. In particular, they studied the transmission of text over an erasure channel, and designed a JSCC scheme using long short term memory (LSTM) networks as encoder and decoder, and showed that this can outperform separation-based approaches using Lempel-Ziv or Huffman coding. A transformer-powered semantic communication system for text, named DeepSC, is proposed in~\cite{Weng:JSAC:21} by utilizing the meaning difference between the transmitted and received sentences. DeepSC is shown to yield better performance than the  traditional communication systems when coping with AWGN channels and it is more robust to channel variations, especially in the low signal-to-noise ratio (SNR) regime.

The core idea behind task-oriented semantic communications for text is to extract the useful information, e.g., grammatical information, word meanings, and logical relationships between words, to achieve intelligent tasks at the receiver,  while ignoring the mathematical expression of words. In~\cite{xie2021task}, Xie~\emph{et al.} have designed a multi-user semantic communication system to execute text-based tasks by transmitting text semantic features. Particularly, a transformer-enabled model, named DeepSC-MT, is proposed to perform the machine translation task for English-to-Chinese and Chinese-to-English by minimizing the meaning difference between sentences. The objective of DeepSC-MT is to map the meaning of the source sentence to the target language, which is achieved by learning the word distribution of the target language. Therefore, cross entropy is utilized as the loss function, represented as
\begin{equation}\label{text1 equation}
{\mathcal L}_{\mathtt M\mathtt T}=\mathbb{E}\left[-\sum_nP\left(\boldsymbol p\lbrack n\rbrack\right)\log\left(P(\widehat{\boldsymbol p}\lbrack n\rbrack)\right)\right],
\end{equation}
where $P\left(\boldsymbol p\lbrack n\rbrack\right)$ and $P(\widehat{\boldsymbol p}\lbrack n\rbrack)$ are the real and predicted probabilities that the $n$-th
word appears in the real translated sentence $\boldsymbol p$ and the predicted translated sentence $\widehat{\boldsymbol p}$, respectively.

Visual question answering task is investigated in~\cite{xie2021task} based on a multi-modal multi-user system. Particularly, the compressed text semantic features and image semantic features are extracted by a text semantic encoder and image semantic encoder at the transmitter, respectively, besides, a layer-wise transformer-enabled model is utilized at the receiver to perform the information query before fusing the image-text information to infer an accurate answer.

\subsubsection{Task-Oriented Semantic Communications for Speech}
The semantic extraction of speech signals is more complicated than text information. For speech signals, the semantic information required for transmission may refer to text information, emotional expression and type of language, etc., which increases the difficulty in extracting semantic features. Weng~\emph{et al.} has investigated a semantic communication system for speech signal reconstruction in~\cite{Weng:JSAC:21}, which aims to minimize the MSE between the input and recovered speech sequences. Moreover, in~\cite{Weng2112:Semantic}, a speech recognition-oriented semantic communication system, named DeepSC-SR, has been developed to obtain the text transcription by transmitting the extracted text-related semantic features. Particularly, two convolutional layers are employed to constrain the input speech signals into a low dimension representation before passing through the multiple gated recurrent unit (GRU)-based bidirectional recurrent neural network (BRNN)~\cite{650093} modules. The text transcription is recognized at the character level by minimizing the difference of the character distribution between the source text sequence and the predicted text sequence. According to the connectionist temporal classification (CTC)~\cite{graves2006connectionist}, the loss function can be expressed as
\begin{equation}\label{CTC loss}
{\mathcal L}_{CTC}(\boldsymbol\theta)=-\ln\left(\sum_{A\in\mathfrak A(\boldsymbol s,\boldsymbol\;\boldsymbol t)}\boldsymbol p\left(\left.\widehat{\boldsymbol t}\right|\boldsymbol s,\boldsymbol\theta\right)\right),
\end{equation}
where $\mathfrak A(\boldsymbol s,\boldsymbol\;\boldsymbol t)$ represents the set of all possible valid alignments of text sequence $\boldsymbol t$ to input speech $\boldsymbol s$, $\boldsymbol p\left(\left.\widehat{\boldsymbol t}\right|\boldsymbol s,\boldsymbol\theta\right)$ denotes the posterior probability to recover one of the valid alignments $\widehat{\boldsymbol t}$ based on $\boldsymbol s$, and $\boldsymbol\theta$ is the trainable parameters of the whole system.

Inspired by DeepSC-SR, a semantic communication system for speech recognition at the word level is proposed in~\cite{han2022semantic}, in which a visual geometry group enabled redundancy removal module is utilized to compress the transmitted data. The objective of this system is to convert the word distribution into the readable text transcription, which is achieved by the cross-entropy loss function between the source word sequence and predicted word sequence. Hyowoon~\emph{et al.} developed a novel stochastic model of semantic-native communication (SNC) for generic tasks~\cite{seo2021semantics}, where the speaker refers to an entity by extracting and transmitting its symbolic representation. 
A curriculum learning framework for goal-oriented task execution is investigated in~\cite{farshbafan2021common}, where the speaker describes the environment observations to enable the receiver to capture efficient semantics based on the defined language by using the concept of beliefs.

\subsubsection{Task-Oriented Semantic Communications for Image and Video}

Due to their high data rate content, significant contribution to the network traffic, and diverse applications from live video streaming to augmented and virtual reality and video gaming, semantic delivery of image and video content is essential for next generation communication systems; and hence, has been studied extensively in the recent years. In \cite{Eirina:TCCN:19}, a neural network aided JSCC scheme was proposed for the first time for efficient delivery of images over wireless channels. The authors proposed an autoencoder-based DeepJSCC scheme, where the channel is treated as an untrainable bottleneck layer. The surprising result in \cite{Eirina:TCCN:19} showed that the proposed DNN-based solution could outperform the concatenation of state-of-the-art image compression techniques (e.g., BPG) with state-of-the-art channel coding (e.g., LDPC) at a prescribed channel SNR. We would like to highlight that DNN-based image compression techniques could only very recently outperform BPG \cite{Balle:NIPS2018}, and their design is quite complex, requiring not only the training of a learned transform coding approach, but also the learning of the distribution of the quantized features for efficient entropy coding. Similarly, so far DNN-based channel code designs cannot meet the performance of state-of-the-art channel codes, such as LDPC, in the long blocklength regime that would be used in image and video transmission. On the other hand, the DeepJSCC scheme proposed in \cite{Eirina:TCCN:19}, and later improved in \cite{Kurka:IZS2020}, can outperform their combination, despite its rather simple architecture. This is because the problem of JSCC is a comparatively easier one for DNNs to learn, since they simply need to learn to map similar signals in the source domain to similar channel inputs, such that after noise addition, they can be mapped to similar reconstructed signals, minimizing the error. On the other hand, learning digital compression and channel coding schemes is a much harder problem due to the structure they need to create, and the discrete nature of the problem makes it more difficult to be learned through SGD. This joint approach also provides significant speed-up in end-to-end delivery. The encoding and decoding tasks can be carried out with significantly less latency in DeepJSCC, thanks to the simple neural network architecture and the complete parallelization it provides, compared to conventional compression and channel coding algorithms, which are often iterative and can be highly complex. 

As mentioned earlier, another significant benefit of the end-to-end DNN-based approach is the graceful degradation it provides; that is, the performance, trained for a specific channel SNR generalizes to other SNRs. This capability was exploited in \cite{Eirina:TCCN:19} to show that DeepJSCC can outperform the separation-based alternatives with even a greater margin when used over a fading channel, when channel state information (CSI) is not available. It is later shown in \cite{Xu:TCSVT:22} that, when the CSI is available at the transmitter and the receiver, an attention mechanism can be used to train a single network that can achieve the best performance at every SNR. 

In \cite{kurka2019successive}, it is shown that DeepJSCC can also achieve successive refinability, that is, the image can be delivered at multiple steps, using gradually more bandwidth, with minimal loss in performance. This means that receivers can tune into the transmission until they recover the transmitted image at the desired quality. \highlight{In \cite{Dai:JSAC:22}, the authors employ adaptive-bandwidth transmission across features depending on feature entropies. This allows allocating more bandwidth to more important features, increasing their reconstruction quality. Bandwidth allocation strategy is learned jointly with the non-linear transform in an end-to-end fashion, significantly improving the performance of DeepJSCC in all performance measures (PSNR, MS-SSIM, LPIPS), especially in the large channel bandwidth regime.} 

\highlight{Another challenge in communication systems is to exploit feedback. It was shown by Shannon that feedback does not increase the channel capacity. Therefore, in the infinite blocklength regime, it does not help from a JSCC perspective either. Since separation is optimal in this regime, what matters for the end-to-end performance is the channel capacity, which the feedback cannot improve. On the other hand, it is known that feedback helps to improve the error exponent in channel coding \cite{Schalkwijk:TIT:66, Gallager:TIT:10}. More interestingly, when transmitting a Gaussian source signal over a Gaussian channel, it is shown in \cite{Schalkwijk:TIT:67} that optimality of uncoded transmission shown in \cite{Goblick:TIT:65} only in the case of matched bandwidth between the source and the channel, extends to arbitrary bandwidth extensions. In \cite{kurka2019deepjsccf}, the DeepJSCC scheme is extended to channels with channel output feedback, and it is shown that feedback can significantly improve the end-to-end performance. It is shown that the required bandwidth for image delivery can be reduced by half when variable rate transmission is allowed and channel feedback is exploited to stop transmission when the required reconstruction quality is reached at the receiver. }

Benefits of DNN-aided JSCC is extended to OFDM channels in \cite{Yang:TCCN:22, Wu:CL:22}. \highlight{In \cite{Wu:CL:22}, the authors employ a double attention mechanism, where one is used to adaptively allocate power according to qualities of the subchannels, a spatial attention mechanism is used to map the features to the subchannels to make sure that the important features are transmitted over the good quality subchannels.} In all these works, the encoder is free to map the input signal to arbitrary points in the channel input space; that is, the channel inputs are limited only by an average power constraint, but there is no input constellation. However, in some practical communication systems, communication hardware is constrained to a fixed constellation diagram. In \cite{Tung:arXiv:22}, a differentiable quantization approach is used to map channel inputs to points from a prescribed discrete constellation, and it is shown that the performance loss compared to DeepJSCC with unconstrained channel inputs is limited as long as a sufficiently rich constellation can be employed. In \cite{choi2019neural}, JSCC of images transmitted over binary symmetric and over binary erasure channels is considered using a variational autoencoder (VAE) assuming a Bernoulli prior. To overcome the challenges imposed by the non-differentiability of discrete latent random variables (i.e., the channel inputs), unbiased low-variance gradient estimation is used, and the model is trained using a lower bound on the mutual information between the images and their binary representations.

\highlight{We would like to highlight that, similarly to neural image compression techniques, an important advantage of DNN-based JSCC approaches, compared to employing conventional compression and channel coding techniques is that, these codes can be trained for any desired final fidelity measure, including various inference tasks, that would not require a complete reconstruction of the source signal. Indeed, it has been shown that DNN-based JSCC approaches outperform their conventional counterparts particularly in terms of SSIM and MS-SSIM performance measure, which are known to better capture the perceived quality, or semantics, of the transmitted images. Similarly, adversarial measures can also be used to further improve the perception quality of the reconstructed images \cite{Yang:TCCN:22}.}

\highlight{A semantic communication system for image retrieval over a wireless channel is considered in \cite{Jankowski:JSAC:21}, which aims to identify the top-$k$ similar images to a query image in a large dataset of images. The authors proposed directly mapping the extracted image features to channel inputs through a three-step training procedure: feature encoder pre-training, followed by JSCC autoencoder pre-training, and finally, end-to-end training. It is shown that this joint approach significantly outperforms the separation-based approach that combines the retrieval-oriented compression scheme mentioned in Section \ref{ss:remote_inference_ML} with a capacity-achieving channel code. This is despite the fact that a very short blocklength is considered, and hence, capacity is far from achievable. This scheme is recently extended in~\cite{xie2021task}. A semantic rate-distortion theory-based communication system for multiple image tasks has been investigated in~\cite{liu2022task}, in which the source image is first restored at the receiver before intelligent task execution. 
}

\highlight{The performance of DeepJSCC for image transmission has also been tested and verified in practical communication systems using a software-defined radio testbed in \cite{Kurka:IZS2020}, which is also exhibited in \cite{Tung:ICASSP:20}. The authors in~\cite{Chae:demo} proposed a real-time semantic testbed based on a visual transformer.
}

\highlight{The first work on the JSCC of video signals over wireless channels employing DNNs is carried out in \cite{Tung:JSAC:22}. In this paper, video signals are divided into group of pictures (GoPs), similarly to common video compression standards. Each GoP is directly mapped to channel input symbols of a fixed bandwidth. The first frame of each GoP is considered as a key frame, and transmitted on its own using JSCC techniques similar to the one used in \cite{Eirina:TCCN:19}. The remaining frames are transmitted using an interpolation encoder, which encodes the motion information in that frame and residual information with respect to the nearest two key frames as reference. Scaled space flow is used to estimate the motion information using the DNN architecture proposed in \cite{Agustsson:CVPR:20}. There are two challenges in the proposed method: First, due to the JSCC for delivering key frames, the encoder does not know their exact reconstruction at the receiver, which depends on the noise realization. The authors use a stochastic encoding method, where the encoder emulates the channel, and generates a reconstruction of the key frame using the channel statistics. The interpolation is based on this stochastically generated version of the key frame. Second, the total bandwidth for the GoP is limited, but in general, one would expect to allocate more channel bandwidth for frames with more motion content. This is achieved by reinforcement learning in \cite{Tung:JSAC:22}. The authors show that the proposed learned bandwidth allocation methodology strictly improves upon equal allocation of the available bandwidth among the frames. The results in \cite{Tung:JSAC:22} show that the proposed JSCC technique for video delivery, called \textit{DeepWiVe}, not only provides graceful degradation with channel SNR, similarly to DeepJSCC, but also outperforms state-of-the-art separation-based digital transmission alternatives combining H.264 or H.265 video encoding with LDPC channel coding at a specified channel SNR value. These results are promising as they show the potential advantages of DNN-based JSCC techniques for future augmented/virtual reality (AR/VR) applications for wireless headsets. 
}

A semantic communication system for image classification has been proposed in~\cite{9606667}, which performs a variational IB (VIB) framework to resolve the difficulty in mutual information computation of the original IB~\cite{TPB99}. The adopted loss function can be expressed as 
\begin{equation}\label{image1 equation}
\begin{split}
    \mathcal{L}_{VIB}\left(\boldsymbol\phi,\boldsymbol\theta\right)&={\mathbf E}_{p\left(\boldsymbol x,\boldsymbol y\right)}\{{\mathbf E}_{p_{\boldsymbol\phi}\left(\widehat{\boldsymbol z}\left|\boldsymbol x\right.\right)}\left[-\log\left(q_{\boldsymbol\theta}\left(\boldsymbol y\left|\widehat{\boldsymbol z}\right.\right)\right)\right] \\
    &+\beta D_{KL}\left(p_{\boldsymbol\phi}\left(\widehat{\boldsymbol z}\left|\mathbf x\right.\right)\parallel q\left(\widehat{\boldsymbol z}\right)\right)\},
\end{split}
\end{equation}
where $\boldsymbol x$ represents the input image, $y$ denotes the target label, and $\widehat{\boldsymbol z}$ is the recovered semantic information at the receiver. $q\left(\widehat{\boldsymbol z}\right)$ and $q_{\boldsymbol\theta}\left(\boldsymbol y\left|\widehat{\boldsymbol z}\right.\right)$ are two variational distributions to approximate the true distributions of $p\left(\widehat{\boldsymbol z}\right)$ and $p_{\boldsymbol\theta}\left(\boldsymbol y\left|\widehat{\boldsymbol z}\right.\right)$, respectively. $\boldsymbol\phi$ and $\boldsymbol\theta$ are the trainable parameters at the transmitter and receiver, respectively. $D_{KL}\left(\cdot\right)$ indicates the Kullback-Leibler divergence. 

\subsection{Distributed Training over Noisy Channels}
We can also extend the model training tasks presented in Section \ref{ss:Compress4Training} for rate-limited channels to training over noisy channels. Since model training is often carried out over many iterations, training among wireless devices imposes strict delay constraints per iteration. Hence, the conventional approach of separate model compression and communication would not meet the desired delay and complexity requirements \cite{Gunduz:CM:20}. 

The problem of training and delivering a DNN to a remote terminal, called AirNet, is considered in \cite{Jankowski:ISIT:22} - extending the model in Fig. \ref{fig:remote_training} by replacing the rate-limited error-free link with a noisy wireless channel. The conventional approach would be to first train a DNN, which is then delivered reliably over the bandwidth-limited channel. We can either train a low complexity model, such as MobileNet or ShuffleNet, or first train a larger model, and then compress it to the level that can be delivered over the limited capacity wireless link. Here, the size of the delivered model will be dictated by the available channel capacity, and errors over the channel will further reduce the accuracy of inference at the decoder side.

An alternative joint training and channel coding approach is considered in \cite{Jankowski:ISIT:22}, where the trained neural network weights are delivered over the wireless channel in an analog fashion; that is, they are mapped directly to the channel inputs. However, given the large size of DNNs, this would require a very large bandwidth. Moreover, the receiver will recover a noisy version of the network weights, which varies according to the noise realization. The authors propose two distinct strategies to remedy these problems. Pruning \cite{l1_pruning} is employed to reduce the network size without sacrificing its performance significantly. The encoder first trains a large-dimensional DNN, which will be then pruned to the available channel bandwidth. Choosing a large DNN as an initial point, rather than directly training a DNN of dimension equal to the available channel bandwidth is motivated by the literature \cite{lottery_ticket_pruning}, which shows that pruning a trained large-dimensional DNN generally performs better than directly training a low-dimensional DNN. The noise problem is remedied by injecting a certain amount of noise to the network's weights at each training iteration, so that the trained network acquires robustness against channel noise. It is shown in \cite{Jankowski:ISIT:22} that the analog transmission of DNN weights achieves better accuracy compared to their digital delivery. A further unequal error protection strategy is also incorporated by pruning the network to a size smaller than the available channel bandwidth, and applying bandwidth expansion to a selected subset of more important weights using Shannon-Kotelnikov mapping \cite{kotelnikov}. 

While the above framework assumes the availability of the dataset at the encoder, in many practical settings, the encoder may simply have access to the DNN architecture and weights but no data. Delivering such a network over a wireless channel is considered in \cite{yulin_dnn}, where it is shown that a Bayesian approach at the receiver when estimating the noisy DNN weights can significantly improve its performance. The authors assume a Gaussian prior, and propose a population compensator and a bias compensator to the minimum mean square error (MMSE) metric.

\highlight{Another common scenario is when the dataset is distributed across many wireless devices, which collaborate to train a common model in a federated manner. When the devices participating in federated learning share a common wireless medium to the parameter server, this is called federated learning at the edge (FEEL). FEEL was studied in \cite{Amiri:TSP:20} and \cite{Amiri:TWC:20} for AWGN and fading channels by first treating the uplink channel from the devices to the parameter server as a MAC, and the devices communicate at the boundary of the capacity region. Then, each device, depending on the rate available to it, reduces the size of its model update by compressing it using the technique proposed in \cite{Sattler:IJCNN:19}. With this approach, there is a trade0-off between the number of devices participating in the model update at each round, and the accuracy of the updates they can convey to the parameter server. The higher the number of devices, less wireless channel resources they are allocated, and the less the accuracy of the updates they transmit to the parameter server. In \cite{Tran:INFOCOM:19}, the authors studied the trade-off between the energy cost of model updates and the latency in FEEL over fading channels. A joint wireless resource allocation problem is formulated in \cite{Chen:TWC:21} for FEEL over fading channels to maximize the convergence rate of the underlying learning process. }

\highlight{In \cite{Amiri:TSP:20, Amiri:TWC:20}, an alternative `analog' transmission approach is proposed for FEEL, by treating the uplink model transmission as a distributed computation problem over a MAC. Inspired by the optimality of uncoded transmission in certain distributed computation and JSCC problems over MAC \cite{Cover:TIT:80, Lapidoth2010}, these papers proposed uncoded and synchronized transmission of local model updates enables, which enables the the parameter server to directly recover the sum of the updates from multiple terminals.} This `over-the-air computation (OAC)' approach has received significant interest in recent years thanks to its bandwidth efficiency \cite{Amiri:TSP:20, Amiri:TWC:20, Zhu:TWC:21, Amiri:TWC:21}. Instead of allocating orthogonal channel resources to the participating devices, they all share the same bandwidth. Unlike in separate model compression and channel coding, the accuracy of the resultant computation benefits from more transmitters as the goal is to recover the sum of their model updates. OAC can also be used in a fully distributed learning scenario \cite{Ozfatura:Globecom:20, Xing:JSAC:21}, where many computations take place in parallel. We refer the reader to \cite{Chen:JSAC:21} for a comprehensive overview of distributed learning techniques over wireless networks.


\subsection{Solutions to the IB Problem} \label{sec:SolutionsIB}
The IB problem detailed in Section \ref{ss:IB} provides a formulation to design mappings $P_{U|S}$ that allow to extract relevant information within the IB relevance--complexity region, i.e, the pairs of achievable $(\Delta,R)$, by solving the IB problem in~\eqref{eq:IBCriteria} for different values of $\beta$. However, in general, this optimization is challenging as it requires computation of mutual information terms. In this section, we describe how, for a fixed parameter $\beta$, the optimal solution $P^{\beta, *}_{U|S}$, or an efficient approximation of it, can be obtained under: (i) known general discrete memoryless distributions and particular distributions, or particular distributions such as Gaussian and binary symmetric sources; and (ii) unknown distributions and only samples are available to design the encoders and decoders.

\begin{figure}[t]
	\centering
	\includegraphics[width=0.5\textwidth]{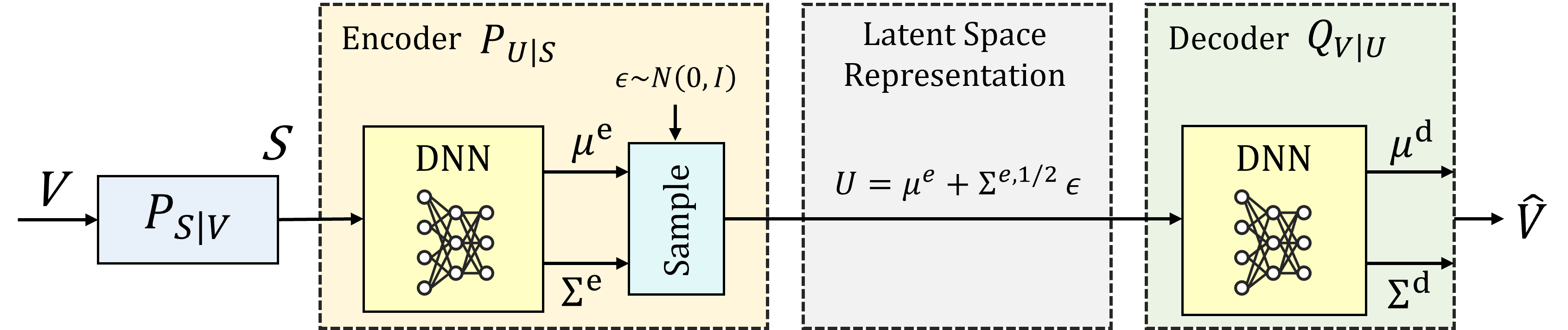}
	\caption{Example parametrization of {Variational Information B}ottleneck  using neural networks.} 
	\label{fig:enc_dec} 
\end{figure}

\subsubsection{Known Discrete Memoryless Distribution}
When the relevant features $V$ and the observation $S$ is discrete and the joint distribution $P_{S,V}$ is known,   the maximizing distributions $P_{U|S}$ in the IB problem in \eqref{eq:IBCriteria}, can be efficiently found by an alternating optimization procedure similar to the expectation-maximization (EM) algorithm \cite{Dempster1977} and the standard Blahut--Arimoto (BA) method~\cite{B72,A72}, which is commonly used in the computation of rate-distortion functions of discrete memoryless sources. In particular, a solution $P_{U|X}$ to the constrained optimization problem is determined by the following self-consistent equations, for all $(u,s,v) \in  \mathcal{U} \times \mathcal{S} \times \mathcal{V}$, ~\cite{TPB99} 
\begin{eqnarray}
P_{U|S}(u|s) = \frac{P_U(u)}{Z(\beta,s)} \exp \Big(-\beta D_{\mathrm{KL}}\Big(P_{V|S}(\cdot|s)\|P_{V|U}(\cdot|u)\Big)\Big) \nonumber\\
P_U(u) = \sum_{s \in  S} P_S(s)P_{U|S}(u|S) ~~~~~~~~~~~~~ \nonumber\\
P_{V|U}(v|u) = \sum_{s \in  S} P_{V|S}(v|s)P_{S|U}(s|u)~~~~~~~~~
\label{self-consistent-equations-IB-algorithm}
\end{eqnarray}
where $P_{S|U}(s|u) = P_{U|S}(u|s)P_S(s)/P_U(u)$ and $Z(\beta,s)$ is a normalization term. It is shown in~\cite{TPB99} that alternating iterations of these equations converges to a solution of the problem for any initial $P_{U|S}$. However, different to the standard {Blahut--Arimoto} algorithm for which convergence to the optimal solution is guaranteed, convergence here may be to a local optimum only.

\subsubsection{Unknown Distributions}\label{ssec:unknow}

The main drawback of the solutions presented in the previous section is that requirement knowing the joint distribution $P_{S,V}$, or at least a good estimation of it and that iterating \eqref{self-consistent-equations-IB-algorithm} can only be performed for sources with small alphabet (or jointly Gaussian~\cite{GT04, CGTW05,  UEZ20}). The Variational IB (VIB) method was proposed in~\cite{AFDM17} as a means to obtain approximate solutions to the IB problem in the case in which the joint distribution is unknown and only a give training set of $N$ samples $\{(s_i,v_i)\}_{i=1}^N$ is available or the alphabet is too large. The VIB consists in defining a variational (lower) bound on the cost of the IB problem in \eqref{eq:IBCriteria}, use neural networks to parameterize this bound and show that, leveraging on the reparametrization trick~\cite{KW13} its optimization can be performed through stochastic gradient descendent (SGD). From a task oriented communication perspective, the VIB approach provides a principled way to generalize the evidence lower bound (ELBO) and Variational Autoencoders \cite{KW13} (and its extension to $\beta$-VAE cost~\cite{HMPBGBML16}) to scenarios in which the decoder is interested in recovering the relevant information $V$ for a task that is not necessarily the observed sample $S$ by maximizing relevance. The idea is to use the IB principle to train an encoder and decoder, parameterized by DNNs, which are able to extract the relevant information to forward to a decoder in charge of reconstructing the relevant information. The resulting architecture to optimize with an encoder, a latent space, and a decoder parameterized by Gaussian distributions is shown in Fig.~\ref{fig:enc_dec}.  This approach has been used for task oriented communications also in JSCC scenarios, as a means to extract the relevant information to transmit over communication noisy channels to perform a given task at the destination, e.g., \cite{shao2021learning},\cite{pezone2022goal}.

\begin{figure}[t]
\centering
\includegraphics[width=0.47\textwidth]{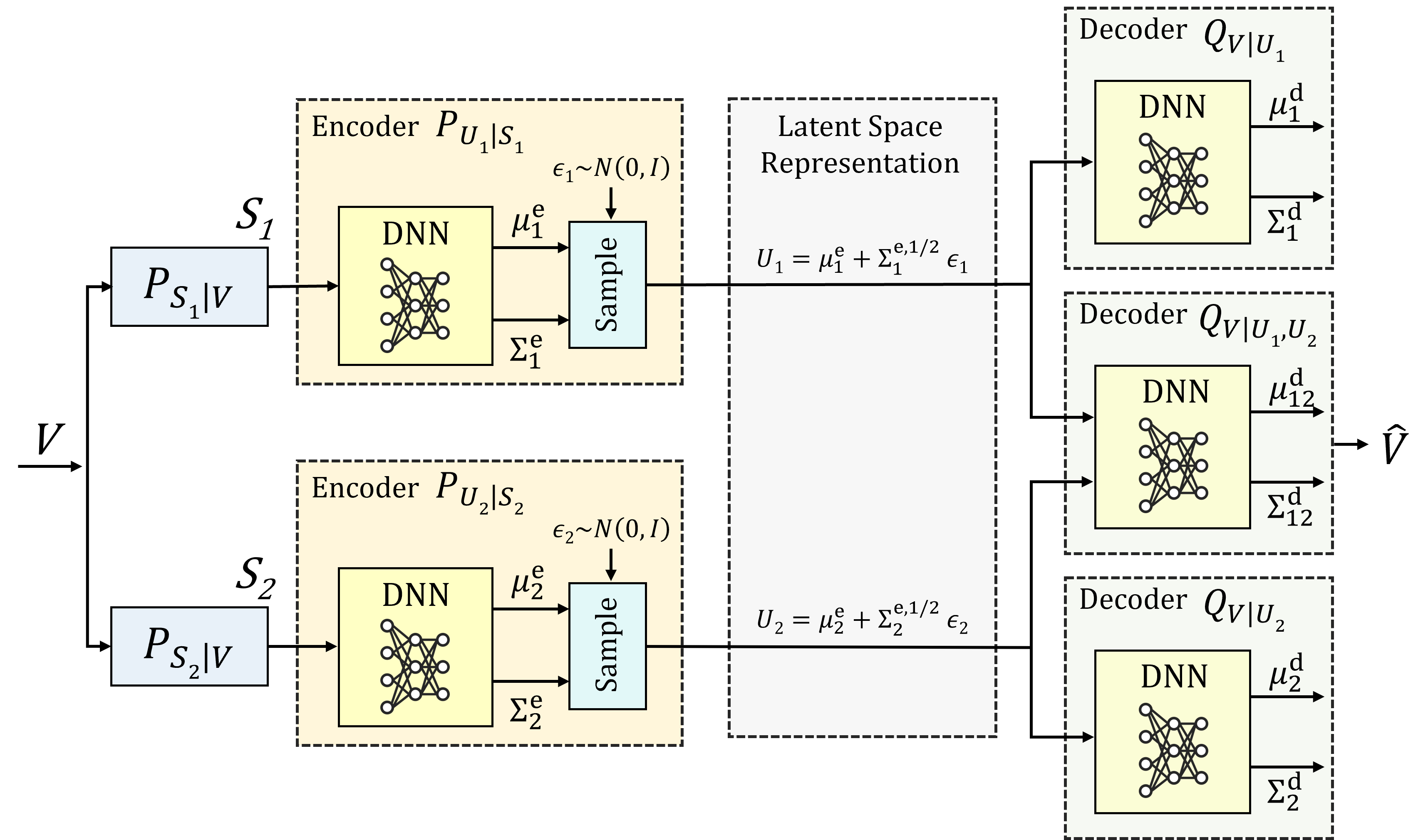}
\caption{Example parameterization of the {distributed variational information b}ottleneck method using neural networks for $K=2$ for encoders parameterized by Gaussian distributions,  $P_{\theta}(u_k|s_k) = \mathcal{N} (s_k; \boldsymbol{\mu}_k^{\theta}(s),  \Sigma_k^{\theta}(s))$.}
\label{fig:latentmodel}
\end{figure}

More precisely, solving the IB problem in \eqref{eq:IBCriteria} consists in optimizing the IB-Lagrangian
\begin{equation}
    \mathcal{L}_{\beta}^{\mathrm{IB}}(P_{U|S}):= I(U;V) - \beta I(U;S).
\label{IB-Lagrangian-formulation-maximizing-relevance-2}
\end{equation}
over all $P_{U|S}$ satisfying $U - S - V$. It follows from Gibbs inequality, for a any $P_{U|S}$ satisfying $U - S - V$, we have the following lower bound on the IB-Lagrangian,
\begin{eqnarray}
\mathcal{L}_{\beta}^{\mathrm{IB}}(P_{U|S})\geq 
\mathcal{L}^{\mathrm{VIB}}_{\beta} (P_{U|S},Q_{V|U},S_{U})  ~~~~~~~~~~~~\\
:=\mathrm{E}_{P_{U|S}}\left[ \log Q_{V|U}(V|U)\right ]  - \beta D_{\mathrm{KL}}(P_{U|S}|S_{U}), \label{variational-bound-IB-problem}
\end{eqnarray}
where $Q_{V|U}(V|u)$ is a given stochastic map $Q_{V|U} \: :\: \mathcal{U} \rightarrow [0,1]$ (also referred to as the variational approximation of $P_{V|U}$ or decoder) and $S_{U}(u):\mathcal{U} \rightarrow [0,1]$ is a given stochastic map (also referred to as the  variational approximation of $P_U$ or prior), and  $D_{\mathrm{KL}}(P_{U|S}|S_{U})$ is the relative entropy between $P_{U|S}$ and $S_U$. The equality holds iff $Q_{V|U}= P_{V|U}$ and $S_{U} = P_U$, i.e., the variational approximations match the true values.

Therefore, optimizing \eqref{IB-Lagrangian-formulation-maximizing-relevance-2} over $P_{U|S}$ is equivalent to optimizing the variational cost \eqref{variational-bound-IB-problem} over $P_{U|S}$,  $Q_{V|U}$ and $S_{U}$.
 In the VIB method, this optimization is done by further parameterizing the encoding and decoding distributions $P_{U|S}$, $Q_{V|U} $, and $S_U$ that are to optimize using a family of distributions  $P_{\theta}(u|s)$, $Q_{\psi}(v|u)$, and $S_{\varphi}(u)$, whose parameters are determined by DNNs with parameters $\theta,\phi$, and $\varphi$ respectively.
 A common example  is the family of multivariate Gaussian distributions \cite{KW13}, which are parameterized by the mean $\boldsymbol{\mu}^{\theta}$ and covariance matrix $\Sigma^{\theta}$.  Given an observation $X$, the values of $(\boldsymbol{\mu}^{\theta}(s), \Sigma^{\theta}(s))$ are determined by the output of the DNN $f_{\theta}$, whose input is $S$, and the corresponding family member is given by $P_{\theta}(u|s) = \mathcal{N} (u; \boldsymbol{\mu}^{\theta}(s), \Sigma^{\theta}(s))$. Another common example are Gumbel-Softmax distibutions \cite{Maddison2016,Jang2017}).
 
The bound \eqref{variational-bound-IB-problem} restricted to family of distributions  $P_{\theta}(u|s)$, $Q_{\psi}(v|u)$, and $S_{\varphi}(u)$ can be approximated using Monte-Carlo and the training samples $\{(s_i,v_i)\}_{i=1}^N$. To facilitate the computation of gradients using backpropagation~\cite{KW13}, the reparameterization trick~\cite{KW13}  is used to sample from $P_{\theta}(U|S)$. In particular, consider $P_{\theta}(U|S)$ to belong to a family of distributions that can be sampled by first sampling a random variable $Z$ with distribution $P_{Z}(z)$, $z\in \mathcal{Z}$ and then transforming the samples using some function $g_{\theta}:\mathcal{S} \times \mathcal{Z} \rightarrow \mathcal{U}$ parameterized by $\theta$, such that $U = g_{\theta}(s,Z)\sim P_{\theta}(U|s)$, e.g. a Gaussian distribution. The reparametrization trick is used to approximate by sampling $M$ independent samples $\{u_{m}\}_{m=1}^M \sim P_{\theta}(u|s_{i})$ for each training sample $(s_{i},v_i)$, $i=1, \ldots, N$. Then, the lower bound on the IB cost can be optimized  using optimization methods such as SGD or ADAM~\cite{Kingma2014} with backpropagation over the  the DNN parameters  ${\theta},\phi,\varphi$ as,
\begin{eqnarray}
 \max_{\boldsymbol{\theta},\boldsymbol{\phi}, \boldsymbol\varphi}\frac{1}{N}\sum_{i=1}^N\mathcal{L}^{\mathrm{emp}}_{\beta,i,M}({\boldsymbol\theta, \boldsymbol\phi, \boldsymbol \varphi} ),\label{eq:Lhatoptimization}
\end{eqnarray}
where the cost for the $i$-th sample in the training dataset is:
\begin{eqnarray}
\mathcal{L}^{\mathrm{emp}}_{\beta,i,M}({\theta, \phi, \varphi} ) \hspace{2in} \label{eq:variational objective2} \\
:=\frac{1}{M}\sum_{m=1}^M\Big[\log Q_{\phi}(v_i| u_{i,m})
-\beta D_{\mathrm{KL}}(P_{\theta}(U_{i}|s_{i})\|Q_{\varphi}(U_{i}))
\Big)\Big],  \nonumber
\end{eqnarray}
and sampling is performed by using $u_{i,m} = g_{\phi}(s_{i},z_{m})$ with $\{z_{m}\}_{m=1}^M$ i.i.d. sampled from $P_{Z}$ for each $(s_{i},v_i)$ pair. 

For inference, let ${\theta}^*,{\phi}^*, \varphi^*$ be the DNN parameters obtained in training by solving \eqref{eq:Lhatoptimization}. Inference for a new observation $S$, the representation $U$  can be obtained by sampling from the encoders $P_{\theta_k^*}(U_k|S_k)$ and a soft estimate of the remote source $Y$ can be inferred by sampling from the decoder $Q_{\phi^*}(V|U)$. Thus, from a task oriented communication perspective, $P_{\theta_k^*}(U_k|S_k)$ is an encoder trained according to the cost \eqref{eq:Lhatoptimization}  to extract the most relevant representation $U$ for inference of $V$, and $Q_{\phi^*}(V|U)$ is a decoder that is trained to reconstruct the relevant information $V$ from the representation $U$ that minimizes the log loss.

Similarly to the steps followed for the variational IB in Section~\ref{ssec:unknow}, encoders and decoder performing on the region \eqref{eq:ComplexityrelevanceFunction} can be computed by deriving a variational bound on and parameterizing encoding and decoding distributions $P_{U_k|S_k}, Q_{U_k|V}$ using a family of distributions whose parameters are determined by DNNs, and optimize it by using the reparameterization trick~\cite{KW13}, Monte Carlo sampling, and stochastic gradient descent (SGD)-type algorithms. The encoders and decoders  parameterized by DNN parameters $\boldsymbol{\theta},\boldsymbol{\phi},\boldsymbol\varphi$ can be optimized according the distributed IB principles by considering the following empirical Monte Carlo approximation:
\begin{eqnarray} 
\max_{\boldsymbol{\theta},\boldsymbol{\phi}, \boldsymbol\varphi}\frac{1}{n}\sum_{i=1}^n\Big[\!\log Q_{\phi_{\mathcal{K}}}(v_i| u_{1,i,j},\ldots, u_{K,i,j} ) \hspace{.5in}\label{eq:DLhatoptimization}
\\
+ s\sum_{k=1}^K \!\Big(\! \log Q_{\phi_{k}}(v_i|u_{k,i,j})\!-\!D_{\mathrm{KL}}(P_{\theta_k}(U_{k,i}|s_{k,i})\|Q_{\varphi_k}(U_{k,i}))
\Big)\Big],  \nonumber
\end{eqnarray}
where  $u_{k,i,j} = g_{\phi_k}(s_{k,i},z_{k,j})$ are samples obtained from the reparameterization trick by sampling  from $K$ random variables $P_{Z_k}$. The resulting architecture is shown in Fig.~\ref{fig:latentmodel}. This architecture generalizes that from autoencoders to the distributed task oriented communication scenario setup with $K$ encoders.


\section{Timing as Semantics in Communications} \label{sec:timing-semantics}

\highlight{In the previous sections, we have considered goal-oriented communications from a `static' perspective; that is, we considered the scenario in which a transmitter observes a certain signal, and wants to deliver it to a receiver under a certain semantic quality measure, which depends on the underlying task, but neither the source statistics nor the semantic quality measure changes over time. Therefore, time did not really factor into our analysis. However, in many practical systems, particularly those involving control of cyber-physical systems, timing of messages can be as critical as their content. A message that arrives too late can be completely useless no matter how reliably it have been recovered.}

In this section, we will deal with a broad class of problems, where the relevance or value of information related to its timing. A specific sub-class of these problems relates to the popular idea of age of information (AoI). However, as we shall argue shortly, AoI is just one piece of the bigger puzzle when it comes to ``timing as semantics'' in communications. For a systematic exposition, we will first summarize key ideas related to the concept of AoI that will be useful in this discussion. We will then introduce a general real-time reconstruction problem using a rate-distortion viewpoint where semantic information is contained in the timing of the source samples. Along the way, we will explore the connection of age metrics to this rate-distortion viewpoint. 

\begin{figure}[t!]
\centering
\includegraphics[width=\columnwidth]{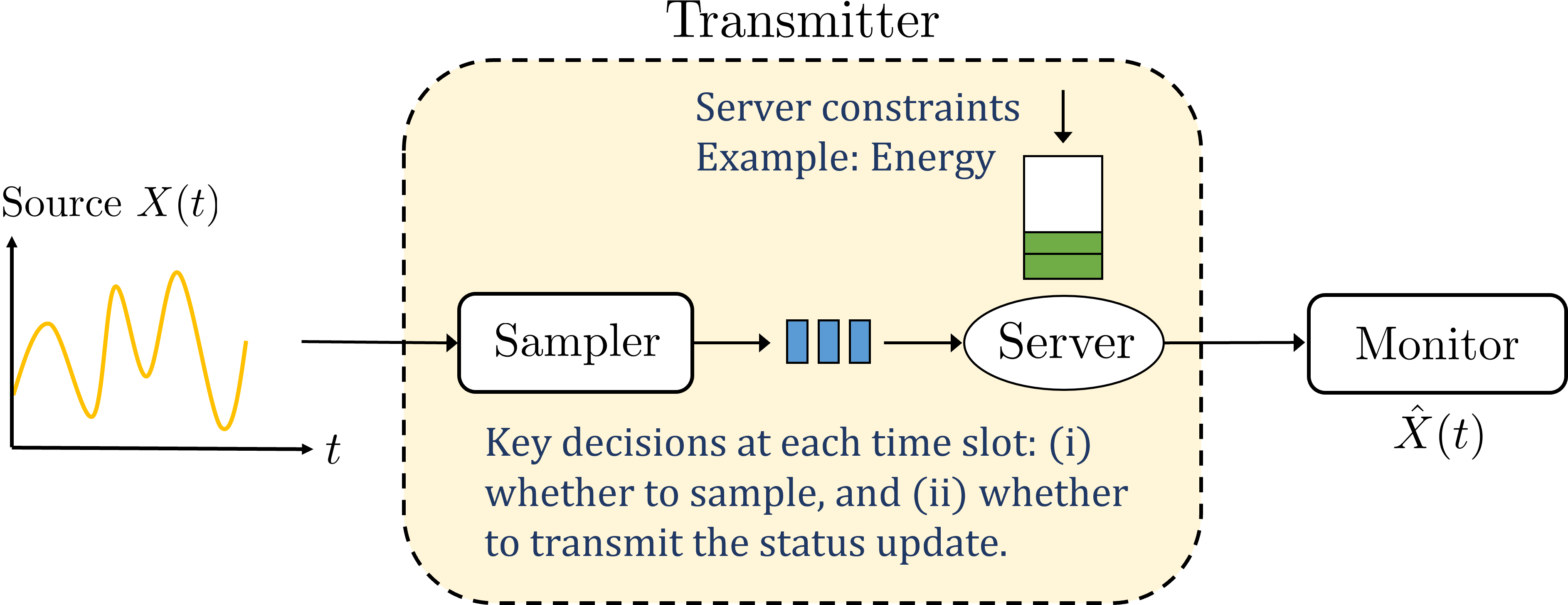}
\caption{An illustration of the general setup for Section~\ref{sec:timing-semantics}.}
\label{fig:hsd:timing-overview}
\end{figure}

The general setup for this section is illustrated in Fig.~\ref{fig:hsd:timing-overview}. It consists of a source modeled as a random process $X(t)$, a transmitter that transmits status updates to a monitor about the current state of $X(t)$, and the monitor where the estimated state is denoted by $\hat{X}(t)$. In this setting, the transmitter needs to take two key decisions during each time slot: (i) whether to sample $X(t)$ to generate a status update, and (ii) whether to transmit the status update. Note that it is not always optimal to sample and transmit in the same time slot when the server is subjected to additional constraints, such as the energy availability/harvesting constraints. With this background, we now present some relevant background on AoI.

\subsection{Age of Information}

\subsubsection{Background and Definition} 
\begin{figure}[t!]
\centering
\includegraphics[width=0.8\columnwidth]{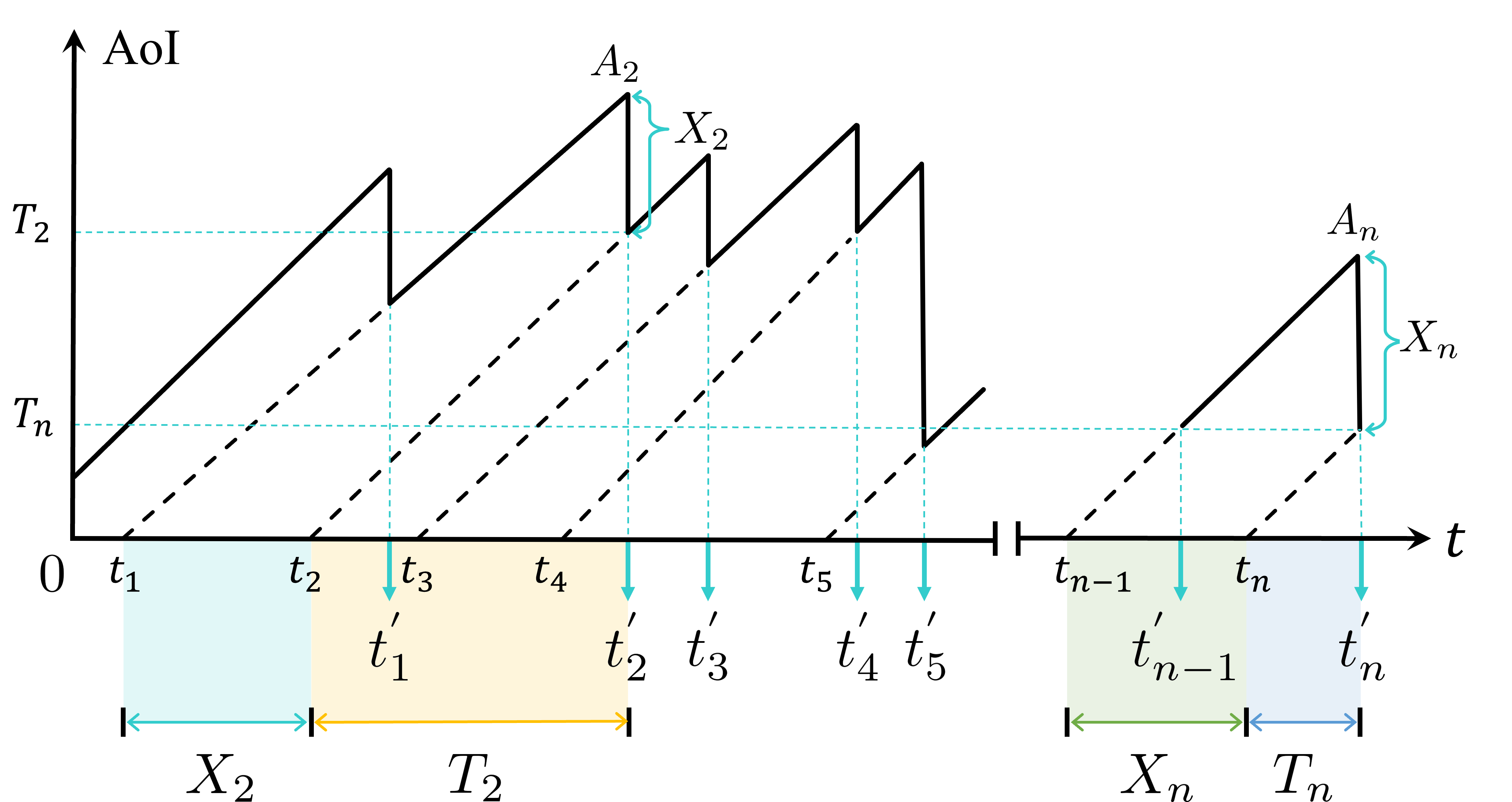}
\caption{AoI evolution vs. time for $n$ update packets~\cite{abd2018role}.}
\label{fig:hsd:illustration}
\end{figure}

AoI is a performance metric that quantifies freshness of information at a monitor about some remote stochastic process $X(t)$ observed by a transmitter node \cite{roy_survey,kosta2017age_mono,abd2018role,sun2019age}. Specifically, AoI is defined as the time elapsed since the last successfully received update packet at the monitor was generated at the transmitter. Mathematically, it is defined as the following random process:
\begin{equation}\label{AoI_def}
\Delta_{\rm AoI}(t) = t - u(t),    
\end{equation}
where $u(t)$ is the generation time instant of the latest status update received at the monitor by time $t$. In order to introduce the idea of AoI concretely, we use Fig. \ref{fig:hsd:illustration}, which depicts a realization of AoI at the monitor as a function of time when the transmitter sends update packets according to a First-Come-First-Served (FCFS) queuing discipline and only one packet transmission may occur at any given time. Here, we implicitly model age as a linear function, which is also the most popular definition of age in the literature, although more general non-linear age functions have also been considered. Further, $t_n$ and $t'_n$ denote the generation and reception time instants of packet $n$ at the transmitter and monitor, respectively. Therefore, we observe that: i) $X_n$ is the inter-arrival time between packets $n - 1$ and $n$, i.e., the time elapsed between the generation of packets $n - 1$ and $n$, ii) $T_n$ is the system time of packet $n$, i.e., the time elapsed from the generation of packet $n$ at the transmitter until it is received at the monitor, and iii) AoI is reset to $T_n$ at $t'_n$ since packet $n$ becomes the latest received update packet at $t'_n$, and hence the AoI value at that time instant is the time passed since the generation of packet $n$, which is $T_n$. Under ergodicity, many key properties of the AoI process, such as the average AoI or the mean peak AoI, can be studied from its sample functions, such as the one depicted in Fig. \ref{fig:hsd:illustration}. Interested readers are advised to refer to well-known books and overview articles on this topic, such as \cite{roy_survey,kosta2017age_mono,abd2018role,sun2019age}, for a more comprehensive introduction to the concept of AoI.

\subsubsection{Age of Incorrect Information (AoII)}
While AoI is useful in characterizing the freshness of information and has found applications in a broad range of system settings under a variety of queuing disciplines \cite{yates2018age,yates2020age,soysal,costa2016age,kam2018age,Moltafet_multisource,kosta2019age,Inoue19,Champati19,abdelmagid_2021a,abdelmagid_2021b}, it does not explicitly take into account the similarity or discrepancy between the {\em status of information} at the transmitter and the monitor. In particular, as observed from \eqref{AoI_def}, the AoI always increases as time passes even if the transmitter and monitor have the same status of information. As an example, consider $X(t)$ to be a discrete Markov source. If the current state of $X(t)$ matches with that of $\hat{X}(t)$, the monitor has the most updated information about $X(t)$ irrespective of when the status update was received. Therefore, in this case AoI is not an accurate measure of information freshness since it ignores semantic information about the nature of source $X(t)$. This motivated the introduction of a related metric, termed AoII, which is defined as \cite{maatouk2020age}: 
\begin{equation}\label{AoII_def}
\Delta_{\rm AoII} = \left(t - v(t)\right) \mathbf{1}\{X(t) \neq \hat{X}(t)\},
\end{equation}
where $v(t)$ is the last time instant when $\mathbf{1}\{X(t) \neq \hat{X}(t)\} = 0$ and $\mathbf{1}(\cdot)$ is an indicator function. Thus, according to \eqref{AoII_def}, AoII grows with time only when $X(t) \neq \hat{X}(t)$. Therefore, AoII explicitly includes some semantic information about the nature of source $X(t)$ that was ignored in the definition of AoI. In order to make this connection more concrete, we now discuss the general real-time reconstruction problem from the rate-distortion viewpoint. 

\subsection{Real-time Reconstruction}

We now establish a connection between the well-known age metrics discussed above and the semantics of information in the context of real-time remote tracking or reconstruction systems as formulated in \cite{Sun:TIT:20}. Recall the setup of Fig.~\ref{fig:hsd:timing-overview}, where a transmitter observes source $X(t)$ and sends samples/updates about that source over time to a monitor. The objective is to reconstruct $X(t)$ at the monitor using the received samples/updates from the transmitter. Recall also that the transmitter has two key decisions to take during each time slot in this general setting: i) whether to sample $X(t)$ or not, and ii) whether to transmit the available or newly generated sample/update or not. Further, we consider a semantic-aware objective function (which is usually referred to as a {\em distortion function}), which can be expressed for the cases of discrete and continuous sources of information as
\begin{equation}\label{obj_discrete}
\mathcal{S}_{\rm dis} = \underset{T \to \infty}{\rm lim}{\frac{1}{T}\sum_{t=1}^{T}{c(t)}},
\end{equation}
\begin{equation}\label{obj_continous}
\mathcal{S}_{\rm con} =  \underset{T \to \infty}{\rm lim}{\frac{1}{T}\int_{t=0}^{T}{c(t) {\rm d}t}},
\end{equation}
where $c(t)$ is an appropriately chosen cost function. For instance, $c(t)$ in the discrete case can be either $c(t) = |X(t) - \hat{X}(t)|$,  $c(t) = a(t) \mathbf{1}\{X(t) \neq \hat{X}(t)\}$, or $c(t) = [X(t) - \hat{X}(t)]^2$, while the latter is also appropriate for the continuous case. 

\highlight{When no channel or transmission delays are considered, the above problem can be thought of as a {\em rate-distortion problem}. In the classical rate-distortion problem, we are interested in the average distortion between the original process at the transmitter and its reconstruction at the receiver under a constraint on the number of bits transmitted. Here, on the other hand, `real-time' operation is considered; that is, the distortion measure does not tolerate delays, and the process needs to be reconstructed at the receiver in a real-time fashion. The general objective here is to develop sampling and transmission policies that minimize the average delay-sensitive distortion. An extreme case is \textit{zero-delay transmission}, where the reconstruction at the receiver at each time depends only on the information received up to that point. Delay-constrained distortion requirement is naturally motivated by cyber-physical applications, where the source signal may model the system state and observations, and the receiver is a controller, which cannot tolerate the infinite delay requirements of the classical rate-distortion formulation.}

\highlight{Specific instances of this problem have already been investigated in the literature. When the transmitter samples and transmits during {\em every} time slot, we obtain a causal compression problem. Real-time compression of a discrete Markov source under a finite rate constraint is studied in \cite{Witsenhausen:BSTJ:79}. In this paper, Witsenhausen showed that if the source is $k$th-order Markov source, then the transmitted codeword at each time instant can depend only on the last $k$ source samples and the present state of the receiver's memory, without loss of optimality. When the source is memoryless, i.e., $k=1$, then it is known that the optimal distortion is achieved by the optimal scalar quantizer (Lloyd-Max) for the source \cite{Gaarder:TIT:82}.}

\highlight{Many works have also considered the sampling aspect in tracking random processes under constraints on the sampling rate. In \cite{Imer:CDC:05, Yonggang:CDC:04, Lipsa:TAC:11}, causal transmission of a discrete random process over a finite time horizon is considered under transmission cost constraints. While \cite{Imer:CDC:05} and \cite{Yonggang:CDC:04} consider special decoder and encoder functions, respectively, and optimize the other one, \cite{Lipsa:TAC:11} made no assumptions on the structure of the encoder or the decoder, and showed that the optimal encoding function is of threshold type and optimal estimation strategy at the decoder is Kalman-like. The optimal distortion-transmission rate trade-off is fully characterized in \cite{Chakravorty:ISIT:15} for Gauss-Markov sources. Continuous-time processes are considered in \cite{Rabi:CDC:06} with a constraint on the number of samples that can be transmitted. Energy, delay and buffer constraints are also considered in \cite{Nayyar:TAC:13, orhan2015source} for delay-limited source delivery.}

\highlight{All of the aforementioned works consider a perfect communication channel, and deal mainly with the rate-distortion performance, rate representing either the usual bits per sample, or the sampling rate. While this can be considered as the extension of the source coding problems overviewed in Section \ref{ss:rate_distortion_theory} with a delay-sensitive distortion measure, in practical systems it is important to take the effect of the channel into account. Note that, since the channel transmission also introduces delays, it will have a direct impact on the achieved distortion.}

\highlight{Optimality conditions for the delay-limited transmission of source samples over a noisy channel have been studied as early as the 60s \cite{Fine:TIT:64, McMillan:BSTJ:69}. These works considered discrete channel inputs and outputs. Real-time transmission of a Markov source over a noisy channel with perfect channel output feedback is studied in \cite{Walrand:TIT:83}. Perfect channel output feedback in this problem significantly simplifies the problem as it allows the encoder to track perfectly the knowledge of the decoder. The more challenging problem of real-time transmission of a Markov source over a noisy channel without feedback is studied in \cite{Mahajan:TIT:09}, and a general framework is introduced to study the corresponding \textit{team decision} problems. A similar JSCC problem with real-time reconstruction constraints is considered in \cite{Gao:CDC:14}, which assumed that the transmitter can only transmit a certain number of times; that is, \cite{Gao:CDC:14} extends the model in \cite{Imer:CDC:05} to the JSCC scenario.}

\highlight{In \cite{Ren:TAC:18}, the authors consider the real-time transmission of a discrete-time Markov source over a fading channel, where each transmission is lost with a probability that depends on the transmission power and the channel state. Assuming that the channel state is known by both the transmitter and the receiver instantaneously, the authors characterize the optimal transmission and estimation strategies for the infinite-horizon average distortion for vector-valued autoregressive sources. They also show that the optimal transmission strategy is of threshold-type when the available transmit power levels are discrete. A similar model is also considered in \cite{Chakravorty:TAC:20}, with the slight but important difference that the channel state is known instantaneously only by the receiver, and is fed back to the transmitter with unit delay. The optimal transmission and estimation schemes are then characterized for Markov and first-order autoregressive sources, and the optimality of threshold based transmission is also shown for discrete power levels.}

\highlight{While these works considered explicit source and channel statistics, and treated the end-to-end optimization problem as JSCC with a delay-constrained distortion measure, there has been a significant research interest from the networking community to treat the physical channel as a random queue \cite{kaul2012real, soysal, hsu2019scheduling, kosta2017age_mono}, and focus on the scheduling aspects rather than source and channel coding. In most of the early works on AoI, the status updates are generated as a new packet at the transmitter by an exogeneous process, and the goal of the transmitter is to choose which packets to transmit to keep the AoI at the receiver low. The core idea here is that dropping some packets can be beneficial to reduce the queue waiting time for future status updates.}

\highlight{While AoI is explicitly defined as the measure of performance in these works, AoI can also be connected to the JSCC framework through a delay-sensitive distortion measure. As mentioned earlier, transmission of a random Wiener process over a random delay channel is considered in \cite{Sun:TIT:20}, and the authors show that the estimation error at the receiver is a function of the AoI. In \cite{Sun:SPAWC:18}, the authors consider the mutual information between the real-time source value and the samples delivered to the receiver, and show that this mutual information depends on the AoI. These provide operational meaning to the AoI measure, and connects it closely with cyber-physical systems and control.}

Next, we present three different sampling and transmission policies \cite{Sun:TIT:20, kountouris2021semantics} that will also highlight subtle connections between the age metrics and semantics of information. 

\subsubsection{AoI-aware Sampling and Transmission Policy} 
In this policy, the transmitter samples a new update and sends it to the receiver once the AoI at the receiver reaches a predefined threshold $A_{\rm th}$. While this policy takes into account the timeliness of the available information at the receiver, it is not semantics-aware since it does not involve the status of information at the source and the receiver in the process of decision-making. This again highlights the fact that AoI is agnostic of the source semantics. 

\subsubsection{Source-aware Sampling and Transmission Policy}
In this policy, the transmitter samples a new update and sends it to the receiver when the status of information at the source changes. This policy can be be thought of as a semantics-aware policy but just from the perspective of the source signal. Since this does not explicitly account for the discrepancy of the status at the transmitter and receiver, it could potentially result in unnecessary transmissions that could have been avoided in a truly semantics-aware policy. One specific example of such an instance given in \cite{kountouris2021semantics} is when a transmission is triggered by a change in the status of the source but the status update does not reach the monitor because of an erasure. Now, if the source returns to the previous state in the next time slot, no transmission is required since there is no discrepancy in the states at the transmitter and receiver. However, the current policy will still transmit an update since it does not incorporate end-to-end semantics. 

\subsubsection{Semantics-aware Sampling and Transmission Policy} 
Not surprisingly, this policy overcomes the drawback of the previous one by accounting for both the status of the source signal and the status of the reconstructed signal at the monitor. In particular, the transmitter samples a new update and sends it to the receiver only if the status of the source signal is different from that of the reconstructed signal. Here, we would like to remind the readers that this discrepancy on the status at the transmitter and the monitor explicitly appeared in the definition of AoII. Therefore, one can also interpret this policy as follows: the transmitter samples a new update and sends it to the receiver at the moment when the AoII becomes non-zero.

\begin{figure}[t]
\centering
\includegraphics[width=0.47\textwidth]{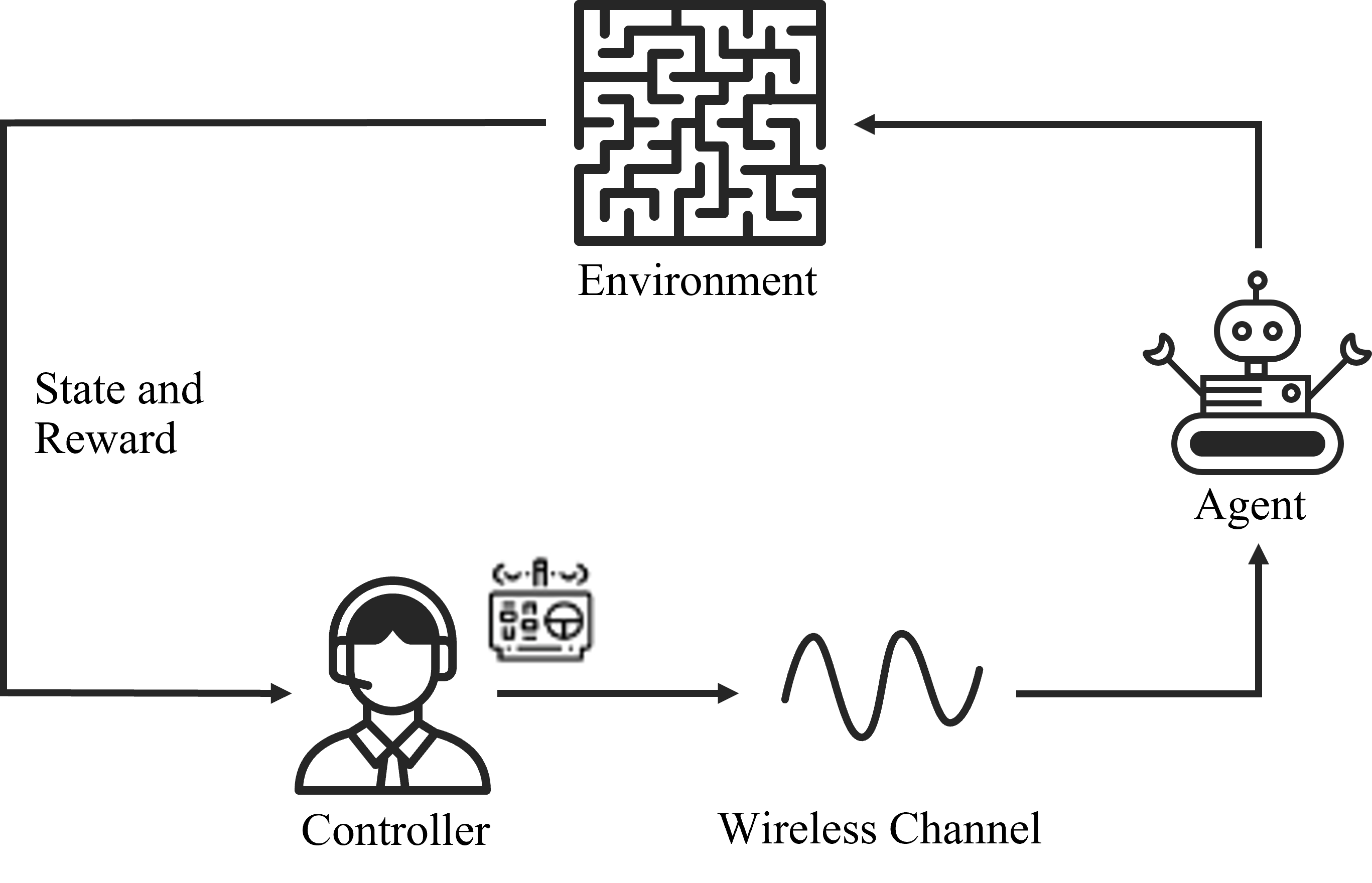}
\caption{Illustration of the effective communication paradigm introduced in \cite{Tung:JSAC:21}.}
\label{fig:remote_control}
\end{figure}

\section{Effective / Pragmatic Communications}\label{sec:Pragmatic}

In \cite{ShannonWeaver49}, Weaver defines Level C of communications as the effectiveness problem, which deals with the effect of the communicated symbols on the receiver. At this level, the communication should have not only a meaning, but also an impact on the behavior of the receiver. In \cite{Tung:JSAC:21}, the authors argue that communication problems where the receiver takes actions based on the messages it receives, and the goal is to maximize a long-term reward function, falls into this category. The reward received following each transmission depends on the state of the receiver as well as the action it takes. In these problems, we cannot define a simple fidelity metric as in the preceding rate-distortion theory based framework. The utility, or fidelity of each transmission depends on its impact over a long time horizon. Here, as in the previous section, time plays an important role, and the same message may have different `meanings' depending on the state of the receiver.  Therefore, we can consider the action taken by the receiver after each transmission as the affect of communication on the receiver, and the state of the receiver as the `context'.

You can see in Fig. \ref{fig:remote_control} an illustration of the effective communication paradigm proposed in \cite{Tung:JSAC:21}.  The authors suggest that any single-agent Markov decision process can be turned into an effective communication problem by considering two entities, one of them, called the \textit{controller}, observes the state and the reward signals, while the actions are taken by the \textit{agent}, which may or may not observe its state, but not the rewards. We further assume that there is a noisy communication channel from the controller to the agent, and the actions taken by the agent will depend on the signals received from the controller through this channel. Note that the controller and the agent collaboratively need to learn the best actions to take to maximize their long term reward, which can be formulated as a reinforcement learning problem. However, unlike in conventional formulation, here the learning must be enabled through the noisy channel.  It is shown in \cite{Tung:JSAC:21} that separating the learning problem from the communication problem will result in a suboptimal performance. 

\highlight{We also note here that both Morris \cite{Morris38} and Weaver \cite{ShannonWeaver49} conceptualized syntax, semantics and pragmatics as going from more specific to more general; that is, pragmatics is the most general of the three. Indeed, the above formulation of effective/pragmatic communication generalizes both the technical communication formulation of Shannon (i.e., the source and channel coding theorems), and the semantic communication as conceptualized in this paper in the form of a JSCC problem with a prescribed distortion measure. To see this, we can formulate the channel coding problem as a two-agent guessing game, where the transmitter observes the message $m \in [M]$ and maps it into a channel codeword $x^n$ that belongs to the set $\mathcal{X}^n$, which denotes the set of possible actions of the transmitter. Upon observing the noisy channel output $Y^n$, the receiver takes one of $M$ actions, $\hat{m} \in [M]$, which correspond to its reconstruction of message $m$. The agents receive a reward of $1$ if $\hat{m} =m$, and $0$ otherwise. Their goal is to maximize their long-term average reward, which corresponds to the probability of error, and their joint policy, which consists of the encoding and decoding functions, form the channel code. Similarly, if we replace the observation of the encoder with $s^m \in \mathcal{S}^m$, its actions with $w \in [M]$, which is observed by the receiver perfectly, the action set of the receiver with $\hat{s}^m \in \hat{\mathcal{S}}^m$, and consider a cost function $d(s^m, \hat{s}^m)$, minimizing the long-term average cost is equivalent to solving the lossy source coding problem.}

\highlight{It is easy to see that the JSCC problem is also a special case of the above effective/pragmatic communication framework, where the encoder observes $s^m \in \mathcal{S}^m$ and takes an action from set $\mathcal{X}^n$, while the decoder observes $\mathcal{Y}^n$ and takes an action from set $\hat{\mathcal{S}}^m$. Minimizing the long-term average distortion, which can be specified as any of the semantic distortion measures mentioned throughout this paper, corresponds to solving the corresponding JSCC problem for the given channel-to-source bandwidth ratio $n/m$. We would like to emphasize that the proposed effective/pragmatic communication framework generalizes semantic communication in multiple directions. First, it can allow memory in the state of the encoder, that is, $s^m(t)$ observed in time slot $t$ can depend on the past state $s^m(t-1)$ or states $s^m(1), \ldots, s^m(t)$. This would result in the type of problems studied in Section \ref{sec:timing-semantics}, in which the goal is to enable the receiver to track a stochastic process available at the transmitter, over a noisy channel. On the other hand, in the model considered in \cite{Tung:JSAC:21}, the actions taken by the receiver also have an impact on the next state of the transmitter.}

It is further shown in \cite{Tung:Asilomar:21} that adjusting the communication scheme according to the state of the agent can result in a better long-term reward, providing evidence for context-dependent communications.

The effective/pragmatic communication framework can be generalized to multiple agent scenarios, where the agents communicate with each other over noisy channels while trying to maximize prescribed reward functions \cite{mostaani_learning-based_2019}. This model generalizes also the recently popular models in the machine learning literature that study the \textit{emergence of communication} among agents within the reinforcement learning literature \cite{foerster_learning_2016, jiang_learning_2018, jaques_social_2019, das_tarmac_2020}. \highlight{In these problems all agents transmit and receive signals in addition to taking actions and interacting with the environment. Communication serves as a tool to enable coordination and cooperation among the agents in order to maximize their accumulated reward, which clearly highlights the pragmatic nature of communication that goes beyond the transmitted bits over the channel or reconstructed signals at the receiver; what matters is not which bits are transmitted, or what the receiver's estimate of the transmitter's signal is, it is the actions taken by the agents.}




\section{Conclusion}
In this paper, we  have presented a comprehensive overview of the foundations of semantic- and task-oriented communications as well as practical data-driven approaches to semantic communication of various information sources including image, text and video. In particular, we argued that semantics can be considered in the context of rate-distortion theory, albeit for most practical information sources it is difficult if not impossible to explicitly state the corresponding distortion metric. Therefore, we provided an extensive overview of deep learning aided approaches to communicating practical information sources over constrained communication channels under various semantic loss functions. In the case of rate-limited error-free communications, we have presented results related to remote inference as well as distributed model training, and exposed the relevance of information theoretic concepts and techniques such as IB. We emphasized the potential benefits of JSCC when communicating over noisy channels, both in terms of end-to-end performance and adaptivity to channel variations.  Finally, we put the time dimension into the framework, and argued that the timeliness of a signal can be considered as yet another aspect of semantics. Finally, we presented an interpretation of the effective/ pragmatic communication problem, which involves communication among agents that take actions to achieve a certain goal. We believe that the provided framework, which focuses on the communications aspects of semantics rather than linguistic formulations, and the in-depth overview of the foundations of semantic- and task-oriented communications will provide guidelines for the development of more efficient practical solutions for semantic- and task-oriented communication networks.

\bibliographystyle{IEEEtran}
\bibliography{references.bib, reference.bib, references_IB, references_hsd}

\end{document}